\documentclass[preprint]{aastex}
\usepackage{lscape,txfonts}

\newcommand{\HI}{\ion{H}{1}}

\newcommand{\ha}{${\rm H}{\alpha}$}
\newcommand{\kms}{km sec$^{-1}$}
\newcommand{\kmskpc}{km sec$^{-1}$ kpc$^{-1}$}
\newcommand{\Msol}{$M_\odot$}

\newcommand{\moyr}{$M_\odot\, {\rm yr}^{-1}$}
\newcommand{\de}{$^{\circ}$}
\newcommand {\mopc}{M_\odot\,{\rm pc^{-2}}}
\newcommand{\gsim}{\lower.7ex\hbox{$\;\stackrel{\textstyle>}{\sim}\;$}}
\newcommand{\lsim}{\lower.7ex\hbox{$\;\stackrel{\textstyle<}{\sim}\;$}}
\newcommand{\pvdiagram}{$p$-$V$ diagram}

\shorttitle{The cold gaseous  halo of NGC 891}
\shortauthors{Oosterloo, Fraternali \& Sancisi.}

\begin{document}

\title{The cold gaseous   halo of NGC 891}


\author{Tom Oosterloo}
\affil{Netherlands Foundation for Research in Astronomy, Dwingeloo, The
Netherlands\\
Kapteyn Institute, Groningen, The Netherlands}
\email{oosterloo@astron.nl}

\author{Filippo Fraternali}
\affil{Department of Astronomy, University of Bologna, Italy
}

\and

\author{Renzo Sancisi}
\affil{INAF - Osservatorio di Bologna, Italy\\
Kapteyn Institute, Groningen, The Netherlands}



\begin{abstract}

We present \HI\ observations of the edge-on galaxy NGC\,891.  These are among
the deepest ever performed on an external galaxy.  They reveal a huge gaseous
halo, much more extended than seen previously and containing almost 30\% of
the \HI. This \HI\ halo shows structures on various scales. On one side, there
is a filament extending (in projection) up to 22 kpc vertically from the
disk. Small ($M_{\rm HI} \gsim 10^6$ \Msol) halo clouds, some with forbidden
(apparently counter-rotating) velocities, are also detected.  The overall
kinematics of the halo gas is characterized by differential rotation lagging
with respect to that of the disk. The lag, more pronounced at small radii,
increases with height from the plane.  There is evidence that a significant
fraction of the halo is due to a galactic fountain.  Accretion
from intergalactic space may also play a role in building up the halo and providing
low angular momentum material needed to account for the observed rotation
lag. The long \HI\ filament and the counter-rotating clouds may be direct
evidence of such accretion.

\end{abstract}


\keywords{galaxies: halos --- galaxies: kinematics and dynamics --- galaxies:
structure --- galaxies: individual (\objectname{NGC 891})}

\section{Introduction}

Recent deep observations of the neutral hydrogen of several nearby spiral
galaxies  indicate that up to about 15\% of the neutral hydrogen of a
spiral galaxy is located in the halo. The best examples are: 
NGC\,891 \citep{swa97}; 
NGC\,2403 \citep{sch00,fra01}; 
NGC\,4559 \citep{bar05};
M31 \citep{wes05}; 
UGC\,7321 \citep{mat03}; 
NGC\,253 \citep{boo05b};
NGC\,5775 \citep{lee01} and NGC\,6946 \citep{boo05a, boo07}. 
Overall, the kinematics of the halo gas is quite regular. The main motion is
differential rotation parallel to the plane.  However, the rotation velocities
in the halo are lower than in the disk \citep{swa97, fra02}. In some cases a
small overall radial inflow is found superposed on the rotation
\citep{fra01}. On small scales  strong vertical motions from and towards
the disk are also observed \citep{boo05a, boo07}.  Halo gas is also detected in
\ha\ \citep[e.g.][]{hoo99, ros03}, with kinematics similar to that of the
neutral gas \citep{ran00,tue00,fra04,hea06a,hea07,kam07}, and in X-rays
\citep[e.g.][]{str04}.

The origin of the gaseous halos is still a matter of debate.  The {\sl
galactic fountain} mechanism \citep{sha76} has received most attention to
date.  In this scheme, gas is pushed into the halo by stellar winds and SN
explosions, mostly in the hot ionized phase. This gas travels through the halo,
eventually cools to neutral and falls back to the disk
\citep{bre80}. 
 There is strong observational evidence  supporting the fountain mechanism,  such as the close correlation 
between the distribution of \ha\ and the high-velocity gas found in NGC 6946 
\citep{boo07}, as well as that the star formation rate
appears to correlate with  the luminosity of the X-ray halo \citep{tue06b} 
and with the radio continuum emission of the halo
\citep{dah06}.  However, the sample of galaxies studied so far is still too
small to be able to make any statement on whether flows related to star
formation are the dominant mechanism or not  in particular for the neutral
halos. Furthermore,  there are indications that an ionized halo only
builds up if the star formation rate is above a critical value \citep{tue06b}
which may pose a problem for the \HI\ halo detected in the super-thin LSB UGC
7321, a galaxy with a low star formation rate \citep{mat03}.  The main
problem for the fountain mechanism comes, however, from the study of the
kinematics. Simple (i.e.\ so-called {\sl ballistic}) models have been unable
to reproduce the kinematics of the ionized gas
\citep{col02,hea06a}. Recently, \citet{fra06} have shown that also the
kinematics of the neutral halo gas cannot be explained by ``pure'' ballistic
galactic fountains. This suggests that other effects, such as the interaction
with a pre-existing hot halo or the accretion from intergalactic space, must
play an important role.  There have also been attempts to model the
extra-planar gas as a stationary medium in hydrostatic equilibrium
\citep{ben02, barn05}, or as a cooling flow accretion in a CDM context
\citep{kau06}, but none of these models is able to reproduce the observations
completely.  It is possible that the halo gas is the result of complex
phenomena involving both internal and external processes.

Understanding the origin and nature of the gaseous halos surrounding spiral
galaxies is important for several reasons.  First, the halo is the region
where material can be exchanged between different parts of the galaxy and this
circulation of gas is fundamental for the galactic life-cycle.  Secondly,
galactic halos are the interface between the galaxy, which is visible and well
studied, and the Intergalactic Medium (IGM), the content and properties of
which remain largely unknown.  CDM cosmological models predict that
most of the baryonic material is currently in the IGM \citep[e.g.][]{whi91,
som06}.  The discovery of the halo gas may provide a new and efficient way to
probe the IGM by studying the
exchange of material between galaxies and their environment.

We believe that the halo gas observed in external galaxies is the analogue of the Intermediate- and
High-Velocity Clouds (IVCs and HVCs) of the Milky Way \citep{wak97}.  The
cloud complexes with anomalous velocities found in galaxies like NGC\,2403
\citep{fra02} have similar masses and velocity deviations
with respect to the disk
as the largest galactic HVCs for which the distances are known
(e.g.\ complex A, \citealt{wak01}).  The total gas mass of the HVCs, if
located in the halo at distances up to a few tens of kpc, would be 
of the order of $10^8$ \Msol, similar to that of the \HI\ found in the halos
of external galaxies (e.g.\ NGC\,891, \citealt{swa97}; NGC\,2403,
\citealt{fra02}).  It is therefore interesting to note that some of the HVCs,
in particular complex C, have been found to have a low metallicity \citep[$Z\sim
0.1-0.3$ $Z_{\odot}$;][]{tri03}.  This may indicate that some of the
HVCs are accretion from the surrounding IGM of ``unprocessed''
material onto the disk of our Galaxy. Such accretion may be necessary to
explain the evolution of disk galaxies \citep{naab06}.  Accretion  onto
disk galaxies may also occur as the merging of small, gas-rich satellites
\citep{vdh88, vdh05}.  Recently, a search has been carried out for the
two largest members of the Local Group other than the Milky Way: M31 and M33
\citep{wes05} and a population of high-velocity clouds at large distances from
these galaxies (about 50 kpc from M31,
\citet{thi04}) has been found.  These clouds have typical masses of a few times
10$^5$ \Msol.  Most of them are thought to be remnants of the
accretion of small companion galaxies, although a small fraction might be
primordial gas clouds \citep{wes05}.

In this paper, we present very deep \HI\ observations of the nearby edge-on
spiral galaxy NGC 891 obtained with the upgraded Westerbork Synthesis Radio
Telescope (WSRT).  NGC\,891 is an Sb/SBb galaxy, one of the best studied
nearby edge-on galaxies.
The disk of NGC\,891 shows
intensive star formation at a rate of $\sim 3.8$ \moyr\ \citep{pop04}.  The
halo region has been studied at various wavelengths and shows a variety of
components from radio continuum emission \citep[e.g.][]{all78} to hot diffuse gas
\citep[e.g.][]{bre94}.  NGC\,891 is considered to be very similar to the Milky
Way with regard to mass and stellar components \citep{vdk84} although it has a
higher star formation rate and significantly stronger radio continuum
emission.  A summary of its physical parameters is given in Table
\ref{t_parameters}.

In the past, NGC\,891 has been studied in \HI\ several times with ever
increasing sensitivity \citep[e.g.][]{san79, rup91, swa97}.  The first
indication of extra-planar material had been reported already by
\citet{san79}, but their favored explanation was that of a flaring outer
disk.  Almost two decades later \citet{bec97} proposed a different
interpretation in terms of a warp of the \HI\ disk along the line of sight.
Finally, more sensitive WSRT observations of NGC\,891 revealed a much more
extended extra-planar component \citep{swa97} and a careful modeling of
the full data cube showed that the most likely explanation was that of
an extended halo component rotating more slowly than the disk. Here we
report the results of \HI\ observations that are a factor 5 more sensitive
 than those of \citet{swa97}.
These observations reveal that the gaseous halo is much more extended than
showed by the previous data.

\section{Observations}

The present observations were obtained with the Westerbork Synthesis Radio
Telescope in the period August - December 2002.  In total, 20 complete 12-hour
observations were performed, using five of the standard array configurations.
The combination of these different configurations gives a regular sampling
of the $uv$ plane from the shortest spacing of 36 m to the longest baseline of
2754 m, with an interval of 36 m.  Care was taken not to use data affected
by solar interference that might have compromised the detection of faint,
extended emission from the halo.  The effective integration time corresponds
to that of 17 complete 12-hour observations.  The observing bandwidth is 10
MHz (corresponding to about 2000 \kms), using 1024 channels (with 2
independent polarizations).  An overview of the observational parameters is
given in table \ref{t_obs}.  The data processing was done using the MIRIAD
package \citep{sault95}.  Before and after each 12-hr
observation a standard calibrator was observed (J2052+365 and 3C147) from
which the spectral response of the telescope was determined.  As is standard
practice with the WSRT, during each 12-hr track no additional (phase)
calibrators were observed to monitor the time variation of system properties.
Instead, the large bandwidth allows to determine these by self-calibration of
the continuum image made from the line-free channels of the data.  This
self-calibration was done using a model of the continuum emission based on the
combination of all observations. An advantage of this approach is
that it also gives an excellent removal of the continuum sources in the line
images, as well as a well-calibrated continuum image (see Sec.\ 3.4).

Three datacubes were made with three spatial resolutions
$23.4^{\prime\prime}\times 16.0^{\prime\prime}, 33.2^{\prime\prime} \times
23.9^{\prime\prime} $ and $69.6^{\prime\prime} \times 58.9^{\prime\prime}$
(see Table \ref{t_cubes}).  The line data were combined and gridded into cubes
of 224 channels 8.2 \kms\ wide to which additional Hanning smoothing was
applied.  This results in a velocity resolution of 16.4 \kms.   As always
with radio observations, the highest-resolution data set has the lowest noise
level (0.09 mJy beam$^{-1}$), the 30- and 60-arcsec data sets have noise levels
of 0.10 and 0.12 mJy beam$^{-1}$ respectively .  The 3-$\sigma$ detection
limits over one resolution element in the three datacubes are $1.3\times
10^{19}$ cm$^{-2}$, $6.8 \times 10^{18}$ cm$^{-2}$ and $1.6 \times 10^{18}$
cm$^{-2}$.  The datacubes were cleaned using the Clark algorithm.  In an
iterative procedure, regions with line emission were identified by smoothing
the data to twice the spatial resolution and selecting a clip level by eye to
define the mask where to clean the data.  This procedure was repeated until
convergence was achieved. 

The \HI\ flux integral is $1.92 \cdot 10^2$ Jy \kms\ corresponding to a \HI\
mass of $4.1\cdot 10^9$ \Msol\footnote{we assume a distance to NGC 891 of $9.5\,$Mpc,
\citep{vdk81}; 1$^\prime$ corresponds to 2.76 kpc}.  The mass derived here is
10\% more than found by \citet{san79} and by \citet{rup91}, but the same as
derived by \citet{brauThilk03}.

\section{Observed properties}

\subsection{Density map}

\begin{figure}
\epsscale{1.0}
\plotone{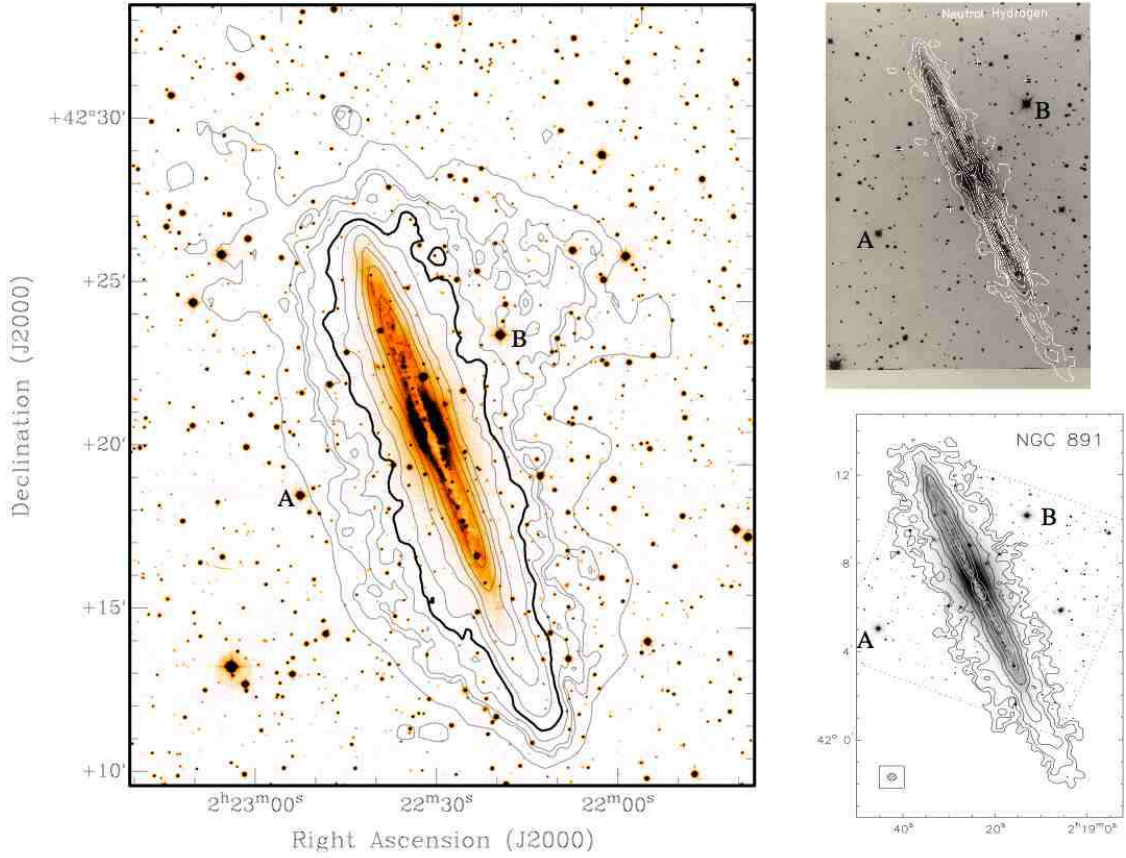}
\caption{ Left: total \HI\ image as obtained from the
observations described in this paper. The outer grey contour comes from the
$60^{\prime\prime}$ data and its level is $0.005\cdot10^{21}$ cm$^{-2}$. The black
contours come from the $30^{\prime\prime}$ data with contour levels 0.01, 0.02,
0.05, 0.1 (thick line), 0.2, 0.5, 1.0, 2.0 and $5.0\cdot 10^{21}$ cm$^{-2}$.
For comparison of the total \HI\ images of NGC 891 published over the
years, we show the one from \citet{san79} (top right) with spatial resolution of
about $30^{\prime\prime}$ and contour levels of 0.5, 1.0, 1.5, 2.4, 4.1, 5.7,
7.3, 8.9 and $10.5\cdot10^{21}$ cm$^{-2}$ and the total \HI\ image as
published by \citet{swa97} (bottom right). The spatial resolution of the last image
is about
20$^{\prime\prime}$ and the contour levels are 0.07, 0.17, 0.46, 1.1, 2.3,
4.1, 6.4 and $9.2\cdot 10^{21}$ cm$^{-2}$. 
To aid the comparison, two stars labeled 'A' and 'B' are marked in each figure 
\label{totalFig}}
\end{figure}

In Fig.\ \ref{totalFig} (right panel) we show the total \HI\ image for
NGC\,891 at $30''$ resolution.  This image was obtained in the standard way by
smoothing the datacube to a resolution of 40$^{\prime\prime}$ and using this
smoothed cube to create a mask to be applied to the original cube.  The mask
was produced by blanking emission below $+3$ r.m.s.\ noise in the smoothed
cube. To illustrate the effects of the improvement in sensitivity of almost
two orders of magnitude, 
in Fig.\ \ref{totalFig} we also give the \HI\ density
distribution as published in two previous studies of NGC 891 \citep{san79,
swa97}.  One feature immediately clear from Fig.\ \ref{totalFig} is that,
despite the large improvement in sensitivity, the radial extent of the \HI\ disk
has {\sl not} become significantly larger and that {\sl in the plane}, the new
observations do not reveal any basically new features (see below).

However, in the vertical direction the situation is dramatically different and
in the new observations the \HI\ extends much farther above the plane.  The
extra-planar emission in NGC\,891, barely visible in the data of \citet{san79}
and detected up to 5 kpc in \citet{swa97}, is now detected up to a projected
distance of more than 10 kpc from the plane everywhere and more than 20 kpc in
a filamentary structure in the NW.  Integrating the flux density located above
and below 1 kpc from the plane, we find that the \HI\ in the halo represents
29\% ($1.2 \times 10^9$ $M_\odot$) of the total \HI\ content of NGC 891.  As
one can see from the lowest contours, the difference in sensitivity between
our total \HI\ map and that of \citet{swa97} is about a factor 5.  The \HI\
observations presented here are among the deepest ever obtained for an
external spiral galaxy.  It is therefore quite possible that also other
galaxies, if observed with comparable high sensitivity, would show similar
extended extra-planar emission and gaseous halos may be a common feature among
spiral galaxies.

\begin{figure}
\epsscale{1.0}
\plotone{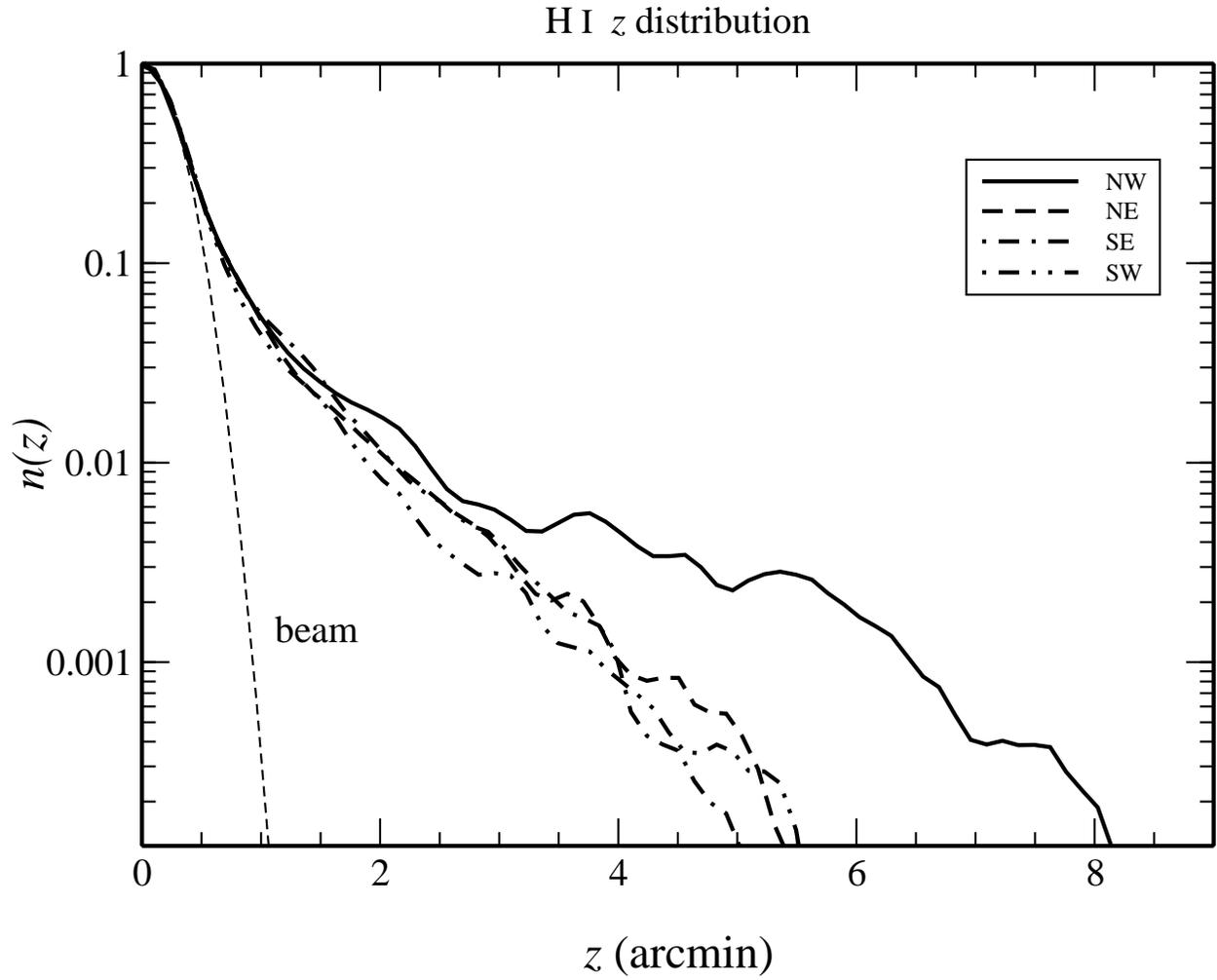}
\caption{Normalized average \HI\ column densities along the $z$ direction for
the four quadrants of NGC\,891. The spatial resolution of the observations 
is indicated. (1$^\prime$ corresponds to 2.76 kpc)
\label{f_zprofs}}
\end{figure}

The vertical extent of the \HI\ layer is illustrated in Fig.\ \ref{f_zprofs}
where we plot the normalised average \HI\ column density, based on the 30
arcsecond dataset, for the four quadrants of the galaxy.  In the NE, SE and SW
quadrants the \HI\ halo extends out to about 5$^\prime$ (14 kpc) while in the NW
quadrant it extends even further out to about 8$^\prime$ ($\sim$22 kpc).  The density
profiles in the NE, SE and SW quadrants are very similar and can be modelled
quite well with an exponential profile with scaleheight of 50 arcsec (2.2
kpc).  In the NW quadrant the density distribution seems to follow the same
trend out to about 3$^\prime$ (8.3 kpc) after which it becomes flatter.  This 
larger extent in the NW quadrant corresponds to a large filament extending
out to more than 20 kpc from the disk (see Fig.\ \ref{totalFig}).

Around the NW filament a crowd of high-latitude clouds are observed  (Figs.\
\ref{f_pvs_minor} and \ref{f_clouds}), which are probably associated with
the filament itself.  However, individual gas clouds are also observed at
large radii and at very anomalous velocities (with large deviations from
rotation).  The middle panels of Fig.\
\ref{f_clouds} show two of these clouds at ``apparently'' counter-rotating
velocities.  The first is in the N-W quadrant at velocities that differ by
about 100 \kms\ from the velocity of the halo at that position.  This cloud is
probably located in the outskirts of the halo, otherwise the drag force of the
halo would invert its motion very quickly.  The second cloud is detected
(Fig.\ \ref{f_clouds}, middle right) at a projected distance of about 28 kpc
from the center of the galaxy.  These clouds have masses of $M_{\rm HI}\sim
1-3 \times 10^6$ \Msol.

As stated above, the new data do not reveal significant new features at large
radii in the plane of the disk. As the earlier data showed, the \HI\ disk of
NGC\,891 is not symmetric, being more extended on the South-West side.  The
fall-off and disappearance of the \HI\ disk on the northern side of NGC 891,
approximately coincident with the end of the stellar disk, and the large
southern extension, confirm the picture of lopsidedness already known from
previous observations and discussed by \citet{bal80}. However, the fact that,
despite the large improvement in sensitivity, NGC 891 does not grow in radius
but does grow substantially in the vertical direction, poses an interesting
question.  The presence and the origin of outer cut-offs of \HI\ disks have
been a matter of debate in the past years.  One of the favored explanations
for their origin has been the effect of the extragalactic radiation field
\citep{mal93}.  In the case of NGC 891 such an explanation would immediately
encounter a serious objection: why would the \HI\ be ionized by a presumably
isotropic radiation field in the plane of the galaxy (and account for the
northern truncation) and apparently not be affected at all in the low-density
halo regions?  Unless the time scales for the disk and for the halo gas and
for their respective ionizations are significantly different, it seems more
likely that the disappearance of \HI\ on the northern side of NGC 891 simply
marks the outer boundary of the gaseous disk.

\subsection{Kinematics}

\begin{figure}
\begin{center}
\epsscale{0.75}
\plotone{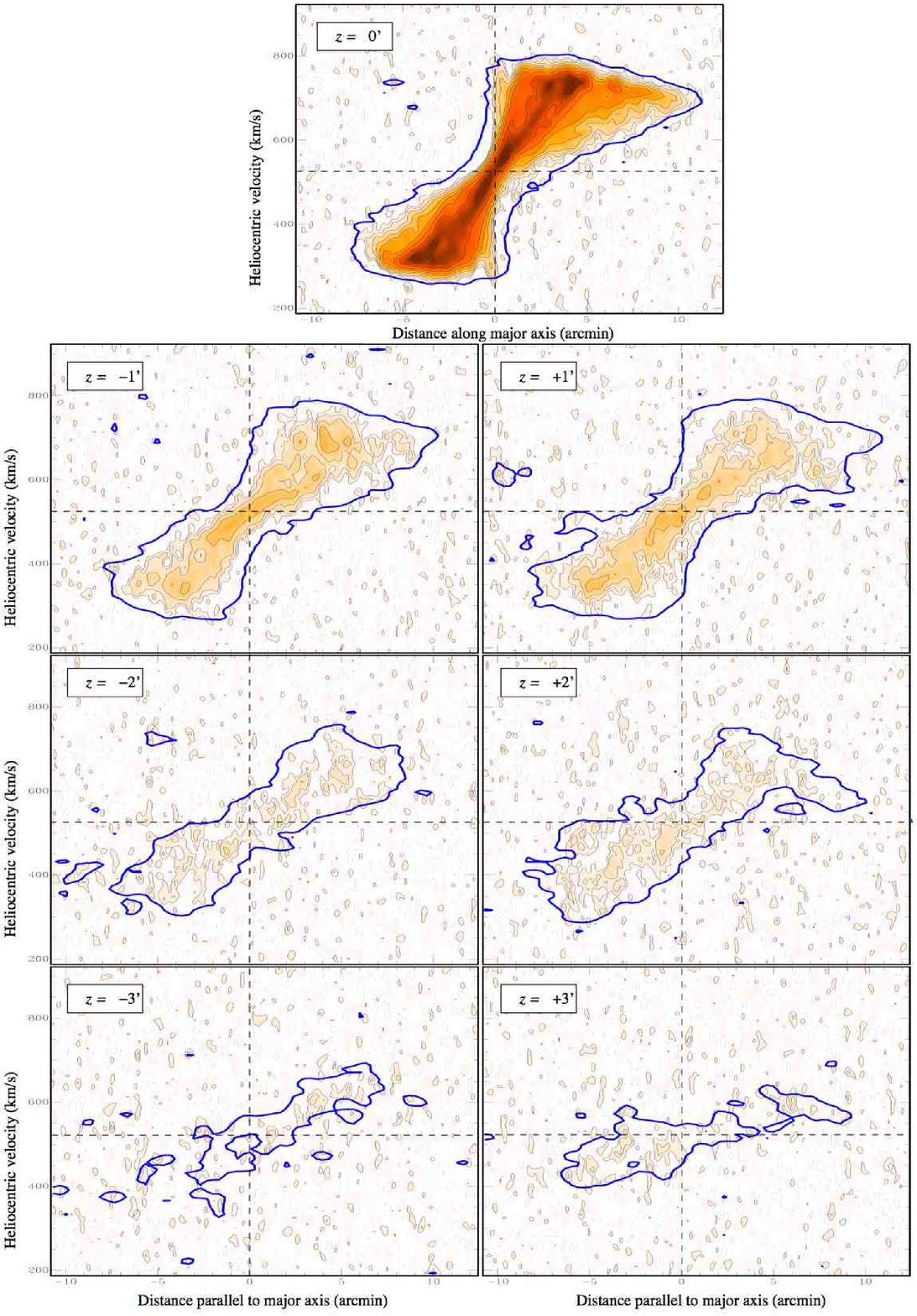}
\end{center}
\caption{Position-velocity plots parallel to the plane for NGC\,891 made from
the high-resolution data cube (thin lines and greyscale) and the 60-arcsec
data (thick line) Contour levels are in steps of 1.5$\sigma$ for the
high-resolution data and  at 3$\sigma$ for the low-resolution data.
\label{f_pvs_major}}
\end{figure}

\begin{figure}
\begin{center}
\epsscale{0.75}
\plotone{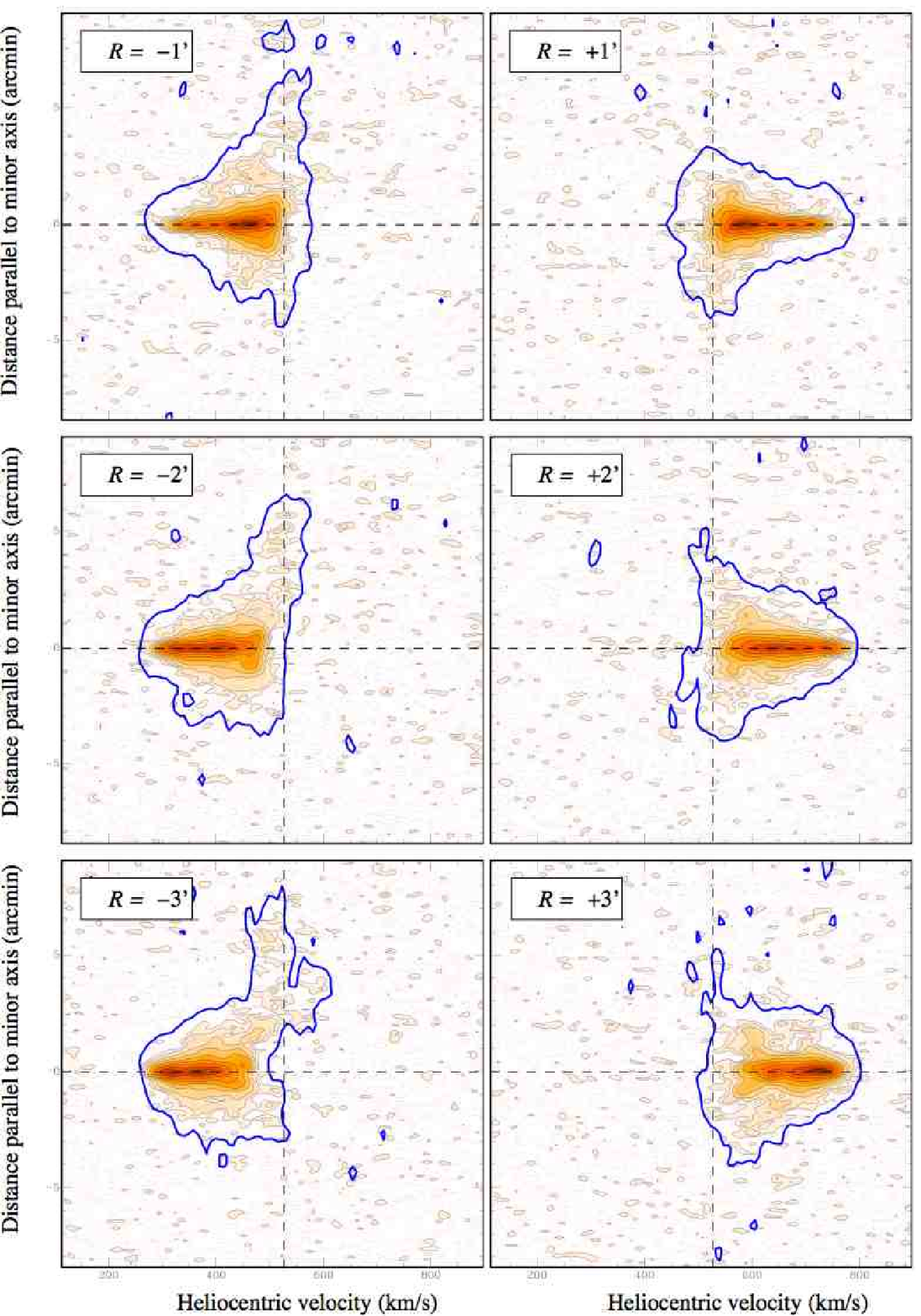}
\end{center}
\caption{Position-velocity plots perpendicular to the plane for NGC\,891  made from
the high-resolution data cube (thin lines and greyscale) and the 60-arcsec
data (thick line). Contour levels are in steps of 1.5$\sigma$ for
the high-resolution data and at 3$\sigma$ for the low-resolution data.
\label{f_pvs_minor}}
\end{figure}

The kinematics of the \HI\ gas is shown by the position-velocity ($p$-$V$)
diagrams parallel and perpendicular to the disk (Figs.\ \ref{f_pvs_major} and
\ref{f_pvs_minor}).  In these diagrams, the high-resolution data are shown with thin contours and
the grayscale, while the thick contours show the low-resolution
(60$^{\prime\prime}$) data at a 3-$\sigma$ level.  The low resolution is
intended to outline the full extent of the faint emission.  The upper plot in
Fig.\ \ref{f_pvs_major} shows the distribution along the major axis (i.e.\ in
the plane) of NGC\,891. The rotation curve is shown by the squares and crosses
in Fig.\ \ref{rotcurve_pv}.  This has been derived by tracing the "envelope',
as already done and described by \citet{san79}, and using a gas velocity
dispersion of 8 km/s.  This method is also used by Fraternali (2007, in
preparation) to derive the rotation curves for the halo gas of NGC 891 at
various distances from the plane.  As already mentioned, the disk of NGC\,891
is lopsided, being more radially extended on the receding S-W side.  The
rotation in the plane is characterized by an inner peak produced by a
fast-rotating inner ring or by a bar \citep{gar95}.  This feature is symmetric
with respect to the center and also with respect to the systemic velocity.
Beyond this, the disk is clearly dominated by differential rotation showing a
roughly flat rotation curve out to a distance of about 6$^\prime$ ($\sim$17
kpc) from the center (Fig.\
\ref{rotcurve_pv}).  The receding side extends further out with an apparent
decrease in rotational velocity.  However, we do not know the azimuthal
location of this extension in the plane of NGC 891.  If this is not along the
line of nodes, the decrease in rotational velocity is only apparent and due to
projection effects.  Note, however, that this extension probably represents a
large fraction of the outer disk as it can be seen in the \pvdiagram\ of Fig.\
\ref{rotcurve_pv} on the entire receding side at projected distances from
0$^\prime$ to 11$^\prime$ and on the velocity side closer to systemic.

Above and below the plane of the disk (lower panels in Fig.\
\ref{f_pvs_major}), the shape of the {\pvdiagram}s changes dramatically.
First, the fast-rotating inner component quickly disappears, indicating that
it is confined to the inner thin disk.  Moreover, the overall shape of the
diagram changes from that of a typical differentially rotating disk to that of
solid-body rotation.  In particular, the two diagrams at $z= \pm 3^\prime$
from the plane clearly show the pattern of a slow solid-body rotator.  These
features are a first indication for a slow rotation of the gas above the plane
with respect to that in the disk.

The {\pvdiagram}s in Fig.\ \ref{f_pvs_minor} show the vertical
density-velocity structure of NGC\,891.  The shape of these plots is generally
triangular with the vertex located at the highest rotation velocity (in the
plane).  This shape was already noted by \citet{san79} in their Figure 7,
showing the \HI\ emission on the receding and the approaching sides.  At
increasing distances upward and downward from the plane, the emission tends to
disappear from the high rotation velocity side and to be restricted closer and
closer to the systemic velocity ($V_{\rm sys}=528$ \kms).  This is
particularly clear for the emission at very high latitudes.  The {\pvdiagram}s
at distances from $R= -1^\prime$ to $R= -3^\prime$ (i.e.\ at 1$^\prime$ and
3$^\prime$ from the center of the galaxy on the N-E side) show emission at low
levels up to about 8$^\prime$ ($\sim$22 kpc) and concentrated
around the systemic velocity of NGC\,891.  This is the filament visible (Fig. \ref{totalFig}) 
on the NW side of the galaxy.

The {\pvdiagram}s also show that, if one compares the halo gas in the
different quadrants, the kinematics of this gas is quite symmetric up to
heights of about 3$^\prime$ ($\sim$8 kpc).  Beyond this, significant
differences exist between the quadrants.  This symmetric kinematics in the
lower halo could indicate that the gaseous halo in this region is in a
equilibrium situation.

A synoptic view of the kinematical structure of the \HI\ gas in the vertical
direction (from the disk to the halo region) is shown in Fig.\ \ref{zMom}
where the thickness of the \HI\ (measured as the first moment of the
$z$-distribution) is given as a function of position along the major axis and
radial velocity.  At the high velocity side (close to rotation) the
$z$-distribution is much thinner than at the lower rotational velocity side as
also seen in Fig.\ \ref{f_pvs_minor}.

\subsection{UGC\,1807}

The present observations show also UGC 1807, a small, gas-rich companion
located at a projected distance of about 30$^\prime$ ($\sim80$ kpc) from
NGC\,891 (Fig.\ \ref{f_companion})  (not discussed in previous papers on NGC 891). 
In the optical, UGC\,1807 
appears as a LSB galaxy oriented almost face-on.  Our \HI\ data show a
regular distribution of
\HI\ with a symmetric velocity field (Fig.\ \ref{f_companion}).  The outer
radius of the \HI\ disk is 4 kpc and the total \HI\ mass corrected for the primary
beam attenuation is $4.5 \times 10^8$ \Msol.  We note that at the position of
UGC\,1807 this correction is quite large, about a factor 6.

The kinematic parameters of UGC 1807 have been derived with a tilted-ring fit
of the velocity field.  We found a systemic velocity of 627
\kms, 100 \kms\ larger than that of NGC\,891,  and an inclination of about 15\de
$\pm$ 5\de.  The rotation curve (Fig.\ \ref{f_companion}) is very regular and
shows a slow rising in the inner regions up to a rotation velocity of about 100
\kms\ at the last measured point. However, because of the low
inclination and its large uncertainty, the
amplitude of the rotation curve is very uncertain: it can be as low as 70 \kms\ 
or as high as 140 \kms.  Using 100 \kms\ and the measured outer radius, we find 
for UGC\,1807 a total mass of $9\times 10^9$ \Msol.

\subsection{Radio Continuum}

Thanks to the broad band and the long integration time of the present
observations we have been able to construct a very deep radio continuum image
(Fig.\ \ref{f_continuum}), similar in quality to the 21-cm continuum image
published by \citet{dah94}.  The spatial resolution is $17\times12$ arcsec,
the RMS noise is 23 $\mu$Jy beam$^{-1}$. Our data do not contain, however, the
information on the polarization of the radio emission.  In Fig.\
\ref{contProfs} we show the normalized vertical brightness distribution of the
radio continuum halo in the four quadrants of NGC 891. As was found by
\citet{dah94}, the vertical brightness distribution of the radio continuum
closely follows an exponential profile with a scaleheight of 25 arcsec
(corresponding to 1.15 kpc) in the NE, SE and SW quadrants and a bit larger
(1.3 kpc) in the NW quadrant. In the SW quadrant the radio continuum halo
seems to be somewhat more extended.  As in the earlier data on NGC 891, there
is also a strong N-S asymmetry of the continuum emission in the disk. Dahlem
et al.\ argue, based on the similarity in the distribution of the H$\alpha$
emission and the radio continuum, that both are the result of outflows from
the disk driven by star formation. In Fig.\ \ref{contProfs} we also indicate
the vertical distribution of the H$\alpha$ emission of the halo, as determined
by Rand \& Heald (priv. comm.).

\section{Models}

Here we investigate the structure and kinematics of the extra-planar \HI\
emission in NGC\,891 by constructing model datacubes and comparing these with
the observations.  We first consider, following
\citet{swa97}, basic models (Section \ref{s_basicmodels}) in which the
extra-planar emission is produced by a strong warp, a flare and a co-rotating thick
disk.  All these models are easily ruled out by the comparison with the data.
Then, in Section \ref{s_halomodels}, we consider more sophisticated models
made of two components (disk + halo) with different kinematics.

\subsection{Basic models} \label{s_basicmodels}

We build model  datacubes using a modified version of the GIPSY 
\citep{gipsy92} program GALMOD.  This program assumes axi-symmetry and
an \HI\ layer made of concentric rings. For each ring we define the \HI\
column density, the thickness of the layer and the geometrical and kinematic
parameters.  The \HI\ disk of NGC\,891 is not symmetric, being more extended
on the SW (receding) side of the galaxy.  In order to avoid the complications
arising from this lopsidedness we construct the models only for the NE
(approaching) side of the galaxy.

We start with a thin disk model.  The \HI\ radial density distribution in 
the plane of the  disk is derived by considering only
the NE side of the galaxy and the region within $|z|<
30^{\prime\prime}$.  The result is plotted in Fig.\
\ref{f_discdensity}.  We approximate the observed  distribution with an
exponential law at large radii and a depression in the inner regions as
described by:
\begin{equation}
\Sigma(R)=\Sigma_{\rm 0} \left( 1 + \frac{R}{R_{*}} \right) ^{\alpha} \exp{(-R/R_{*})},
\end{equation}
where $\Sigma_{\rm 0}$ is the central  surface density, $\alpha$ is an
exponent defining the tapering towards the center and
$R_*$ is a scale radius such that the peak of the distribution is located at
$R=R_*(\alpha-1)$.  The parameters of the fit shown in Fig.\
\ref{f_discdensity} are $\Sigma_{\rm 0}=6.2 \times 10^{-4}\ \mopc$,
$\alpha=7.8$ and $R_*=1.2$ kpc.  In the inner regions of the galaxy
($R\lsim$3 kpc) this model fails to reproduce the inner ring that is
clearly visible in Fig.\ \ref{rotcurve_pv}.  To reproduce also this
inner ring, we add an extra exponential component to our model.  The
parameters of this component are $\Sigma_{\rm 0}=6.3\ \mopc$ and $R_{*}
= 1.2$ kpc. 
The approaching and receding rotation curves in the disk of NGC\,891 are shown
in Fig.\ \ref{rotcurve_pv}.  In our models, we use the curve derived for
the N-E (approaching) side of the galaxy and we assume a velocity dispersion
$\sigma_{\rm gas}=8$ \kms.

NGC\,891 has a mild warping of the outer disk, as can be seen in the top-right
plot in Fig.\ \ref{f_models_chan1}.  In all models, we have included this mild
warp by varying the inclination and position angles of the rings beyond $R=13$
kpc.  The warping angles that fit the data are quite small (1.5$^\circ$ in
p.a.\ and $4^\circ$ in inclination).  The inclination angle of the inner disk
is taken to be 90$^\circ$.  The vertical distribution has been modeled with an
exponential profile with scaleheight $h_{\rm disk}=0.2$ kpc.

Fig.\ \ref{f_models_chan1} (leftmost column) shows the best fit thin disk
model for NGC\,891.  The rightmost column shows the data channel maps at full
resolution.  An important property of the data is that the {\sl apparent}
vertical thickness increases from very thin in the $v_{\rm hel}=292$ \kms
channel (corresponding to the extreme rotation velocity) to a broad vertical
distribution in the bottom channel (close to the systemic velocity, $v_{\rm
sys}=528$ \kms).  This is a major constraint for all our models as it
indicates that there is no halo gas with rotation velocities as high as those
of the gas in the disk.  Clearly the thin disk model only reproduces the top
channel at $v_{\rm hel}=292$ \kms.  The gas visible in this channel is likely
to be at the line of nodes and its thickness is a measure of the intrinsic
thickness of the thin disk.  However, because of the insufficient resolution
(${\rm HPBW} = 19^{\prime\prime} \simeq 0.9$ kpc in the vertical direction)
this channel gives only an upper limit to the thickness of the thin disk (see
below).

The second column of Fig.\ \ref{f_models_chan1} shows the effect of a strong
warp along the line of sight in addition to the mild warp described above.
Such a strong warp model has been proposed by \citet{bec97} to explain the
extra-planar emission observed in NGC\,891 with the VLA \citep{rup91}, in
contrast with the model of a thick, lagging layer favored by \citet{swa97}.
In the present model we have introduced a linear change in the inclination of
the outer rings (beyond $R=13$ kpc) that reaches a tilt of $25^\circ$ in the
outermost ring with respect to the inner parts ($\rm i =90^\circ$).  This
model comes close for the top and the bottom channels of Fig.\
\ref{f_models_chan1} but totally fails to reproduce the middle ones.  The
characteristic feature of any line-of-sight warp model is the ``butterfly''
opening visible in the middle channels (around \ $v_{\rm hel}=366.2$ \kms)
\citep[see also][]{gen03}. Such a pattern is totally absent in
the present data and, therefore, we can rule out that the
extra-planar emission in NGC\,891 is produced by a line-of-sight warp.

The middle column of Fig.\ \ref{f_models_chan1} shows the effect of a large
flaring of the \HI\ layer.  In this model, the thickness of the \HI\ layer in
the inner regions is as small as  in the thin-disk model ($h_{\rm disk}
\simeq 0.2$ kpc) but it increases outwards and reaches $h_{\rm disk}
\simeq 3$ kpc in the outer rings.  Such an unrealistically large value is
necessary to produce the emission  observed at high latitudes.
This model generates some channel maps fairly similar to those
observed (e.g.\ at $v_{\rm hel}=366.2$ \kms), but it completely fails to
reproduce the thin structure in the top channel maps. This is because in
the model the gas in the outer flare rotates as fast as the gas in the plane.
Also this model, therefore, can be  ruled out.
As already mentioned above, from the thin structure observed at the highest
rotational velocities (top Fig.\ \ref{f_models_chan1}) it is possible to derive an upper limit
to the thickness of the thin disk and to set an
upper limit to its flaring.  We find that the inner scaleheight of the disk is
less than $h_{\rm disk}<0.3$ kpc, while in its outer regions the
disk can flare up to at most $h_{\rm disk} \sim 0.5$ kpc.

We now consider models made of two components: 1) a thin disk (like
that in the leftmost column in Fig.\ \ref{f_models_chan1}) and 2) a thicker
layer or halo.
For the vertical gas distribution of the latter
we use  a density profile described by the
following empirical function:
\begin{equation}
\zeta(z)=
\zeta_{\rm 0}
\frac{\sinh(z/ h_{\rm halo})}{\cosh(z/ h_{\rm halo})^2}
\end{equation}
where $z$ is the vertical coordinate, $\zeta_{\rm 0}$ is the surface density
in the plane and $h_{\rm halo}$ is the halo scaleheight.  With this
formulation, the vertical density of the thick layer is zero in the plane,
then rises reaching its maximum at $z=0.88 h_{\rm halo}$ and declines nearly
exponentially further out.  The half width at half maximum (HWHM) of this
distribution (defined as the $z$ where $\zeta(z)=\zeta(0.88 h_{\rm halo}/2)$
is at $z=2 h_{\rm halo}$. Between disk and halo there is a gradual transition.
In this way the spatial coexistence between the two components is minimized:
the halo takes over when the disk is fading out.

In order to model the gas density in the halo, we have extracted and
de-projected the radial distributions at various heights.  The left panel of
Fig.\ \ref{f_halodensity} shows (squares) two of these deprojected radial
distributions at$z=2.8$ and $z=5.6$ kpc (average of the NE and NW quadrants).
The shapes of these distributions change with the height from the plane: as
the distance from the plane increases, the radial distribution becomes flatter
and less concentrated to the center (cf.\ Fig.\ \ref{f_discdensity}).  We have
fitted the data with the function:
\begin{equation}
\rho_{\rm halo}(R,z) = \Sigma(R) \frac{\zeta(z;h_{\rm halo}(R))}{\zeta_0}
\end{equation}
where $\Sigma(R)$ and $\zeta(z;h_{\rm halo})$ are given by equations 1 and 2
but the scaleheight $h_{\rm halo}$ varies with $R$.  Fig.\ \ref{f_halodensity}
(right) shows the fitted values (squares) of $h_{\rm halo}$ as a function of
$R$ and a power law fit (solid line).  With this parametrization the
scaleheight of the halo varies from $h_{\rm halo}=1.25$ kpc (HWHM$=2.5$ kpc)
in the central regions to $h_{\rm halo} \sim 2.5$ kpc (HWHM$\sim 5$ kpc) in
the outer parts.  The other parameters of the fit are: $\Sigma_{\rm 0}=1.4
\times 10^{-1}\ \mopc$, $\alpha=4.5$, $R_*=1.9$ kpc.  The final result is
shown (lines) in the left panel of Fig.\ \ref{f_halodensity} (at $z=2.8$ and
$z=5.6$ kpc from the plane).

The forth column in Fig.\ \ref{f_models_chan1}  shows a two-component
model in which the halo co-rotates with the disk.  The mass of the halo
component is  $M_{\rm halo}=1.25 \times 10^9$ \Msol, 30\%  of the total \HI\ mass.
Clearly, this model does not
correctly reproduce the channel maps at high rotation velocities (top two
panels) which appear much thicker in the model than in the data.

\subsection{Lagging-halo models} \label{s_halomodels}

 The failure of the two-component model with a co-rotating halo to reproduce
the channels near the extreme rotation velocity suggests that the gas above the
plane is rotating more slowly than that in the plane, i.e.\ the halo is
lagging in rotation with respect to the disk.  Evidence for such a lag was
found by \citet{swa97} and is also  observed in
the ionized gas \citep{hea06b,kam07}. 
Fig.\ \ref{f_models_chan2} shows  four models with lagging halos where
the rotation velocity decreases with increasing height above the disk.  All
these models consist of a thin disk (see Fig.\ \ref{f_models_chan1}) and a
halo component with the density distribution described above.  Table
\ref{t_models} lists the kinematic parameters used in the lagging-halo
models.  The first model (leftmost column) is that of a lagging halo with 
the vertical gradient in rotation velocity independent of radius.  
The rotation curve in the disk is that  shown in Fig.\
\ref{rotcurve_pv} (neglecting the inner fast-rotating ring).  We have adopted,
after a few trials, a constant negative vertical gradient in rotation velocity
of $\Delta v_{\rm rot}/\Delta z=-0.55$
\kms\ arcsec$^{-1}$ $\simeq -12$ \kms\ kpc$^{-1}$ for the halo component.

It is clear that this simple lagging-halo model reproduces
the main features of the data much better than the previous models.
The  structure in the upper channel maps  is as thin as in the data and 
it becomes thicker as one approaches the systemic velocity (bottom row).

However, this model is still not completely satisfactory.  In
particular, near the systemic velocity, the radial extent of the halo gas
is much narrower than in the data. One way to improve the model is to
increase the velocity dispersion of the halo gas (second column of Fig.\
\ref{f_models_chan2}). For  this  model all values of the parameters have been 
kept the same as in the previous model, except the velocity
dispersion of the halo gas which has been increased to $\sigma_{\rm halo} = 25$
\kms\ and the halo rotation which has been decreased.
The higher velocity dispersion is physically plausible since the gas in the
halo is expected to be kinematically ``hotter'' and to have a more complex
motion than the gas in the plane.  However, to such increasing velocity
dispersion corresponds a decreasing halo rotation and, therefore, an
increasing vertical velocity gradient to $\Delta v_{\rm rot}/\Delta z=-0.8$
\kms\ arcsec$^{-1}$ $ \simeq -17.4 $ \kms\ kpc$^{-1}$.  Fig.\
\ref{f_models_chan2} shows that increasing the velocity dispersion of the halo
gas indeed  improves the match of the radial extent of the halo \HI\ in the
channel maps close to the systemic velocity (bottom row).

 An alternative approach to the increase of the velocity dispersion is
to introduce a systematic non-circular motion in addition to the
rotation of the halo gas. Such non-circular motions have been found in
similar studies of other galaxies. Large-scale inflows toward the galaxy
center have been discovered in NGC\,2403  ($v_{\rm rad} \sim
-15$ \kms) \citep{fra01} and in NGC 4559
\citep{bar05}.  The third column of Fig.\ \ref{f_models_chan2} shows the
effect of an overall radial motion of the halo of $|v_{\rm rad}|\lsim 25$ \kms.
This is of about the same amplitude as found in NGC\,2403 (although in
an edge-on galaxy one cannot discriminate between in- and outflow). The
models also show that the effects of a higher velocity dispersion or radial
motions in the halo are very similar.  It is, therefore, not possible to
discriminate between an in/outflow and a higher velocity dispersion of the
halo gas.

All the models described above do reproduce most of the features present in the
data, but do not fully account for the shape of the middle channel maps of
NGC\,891.  Let us focus in particular on the channel map at $v_{\rm hel}=366.2$
\kms. The shape in the data channel map is roughly triangular, while in the models
it is more boxy. Clearly, in the model channels near $v_{\rm hel} = 366.2$
\kms, there is too much halo emission near the center.  In order to obtain
the triangular shape, we consider two possibilities. The first is that the
inner regions of the halo are depleted of gas (much more than shown in Fig.\
\ref{f_discdensity}). This obviously would decrease the amount of gas near the
center in the models. The second possibility is that the gradient in
rotation velocity in the inner parts of the galaxy is larger than in the outer
parts. First we consider the possibility of a stronger
central gas depletion in the halo.  Since the gas density in the halo was
derived from the data (without any assumptions about the kinematics), and
modeled accordingly, we do not expect the used \HI\ density to be
significantly different from the actual one.  The main source of errors is the
de-projection of the radial profiles.  However, this is not expected to have a
strong effect beyond $R>3$ kpc, while in order to reproduce the channel map at
$v_{\rm hel}=366.2$ \kms\ the gas density should be significantly lower than
the one used here for radii as large as $R \sim 8$ kpc.  Therefore, it
seems unlikely that a central depletion is the explanation.

Consider instead the second possibility. The forth column from left in Fig.\
\ref{f_models_chan2} shows a model (named {\sl shallow rise})
in which the vertical gradient in the rotation
velocity varies with $R$.  In the inner regions, the gradient is quite large
($\Delta v_{\rm rot}/\Delta z= -43$ \kms\ kpc$^{-1}$ at $R=0$), and it
decreases linearly with $R$ ($\Delta \left( \Delta v_{\rm rot}/\Delta
z\right)/\Delta R
\sim 2.5$ \kms\  kpc$^{-2}$) until it reaches a value of $\Delta v_{\rm rot}/\Delta
z= -14$ \kms\ kpc$^{-1}$ and remains constant further out. The shape of
the rotation curve of the halo changes with distance from the plane: its inner
rising part becomes shallower and shallower. This
behavior is illustrated in Fig.\ \ref{rotcurves_z}.  Clearly, this model
gives a better representation of the channel at $v_{\rm hel}=366.2$ \kms\
while keeping almost unchanged the other channel maps (note that this shallow-rise
model has also a high $\sigma$ for the halo gas and an inflow similar
to the previous two models).
The rotation curves for the halo gas at 
various distances from the plane, derived directly from the observations,
will be presented and discussed by Fraternali (2007, in preparation).

We further compare  the models by inspecting the position-velocity
cuts perpendicular to the major axis of the galaxy.  Fig.\
\ref{f_models_min2} shows two  sets of such cuts taken on the NE
side of the galaxy at a distance of 1$^\prime$ (2.8 kpc, bottom) and
2.7$^\prime$ (7.5 kpc, top) from the center.  The emission shows a
characteristic triangular shape.  The thin-disk is visible at all
velocities between systemic ($v_{\rm hel}=528$ \kms) and the maximum rotation
($v_{\rm hel} \sim 290$ \kms), whereas the halo appears to have its maximal
extent near the systemic velocity.  It is clear that the shallow-rise model best
reproduces the triangular shape as well as the difference between the
$1^\prime$ and $2.7^\prime$ plots. The white squares in the data plots
show the rotational velocities adopted for the shallow-rise model.

The comparison of  the various models leads to the conclusions that the
thin disk of NGC\,891 is surrounded by an extended gaseous halo which is
rotating more slowly than the disk and contains almost 30\%
of the neutral gas of the galaxy. 
The kinematics of the gas in the halo can be best
explained by assuming a vertical gradient in the rotation velocity. Moreover,
this gradient is stronger in the inner regions ($\Delta v_{\rm rot}/\Delta z
\simeq -43$ \kms\ kpc$^{-1}$) than at larger radii ($\Delta v_{\rm
rot}/\Delta z \simeq -14$ \kms\ kpc$^{-1}$. Finally, the velocity dispersion
of the halo gas is higher than that of the gas in the disk ($\sigma_{\rm
halo}\simeq 20-25$ \kms\ vs $\sigma_{\rm disk} \simeq 8$ \kms) and/or there are
significant radial motions in the halo gas ($|{v_{\rm rad}|} \lsim 
25$ \kms).

\section{Discussion}

In the previous Section we have studied with 3-D models the extended
extra-planar \HI\ emission of NGC\,891 and concluded that this galaxy has
a massive halo of neutral gas rotating more slowly than the disk.
Here we discuss the properties of this \HI\ halo, its possible
origins and the comparison with data at different wavelengths.

\subsection{The structure of the \HI\ halo}

The \HI\ halo of NGC\,891 is the most extended and massive of those found to
date in a spiral galaxy \citep{fra07a}.  The \HI\ mass above $z=1$ kpc is
about 30\% of the total \HI\ mass.  
The distribution of gas in the halo appears fairly symmetric and regular in
the four quadrants of the galaxy up to about 8 kpc from the plane (Figs.\
\ref{totalFig} and \ref{f_zprofs}).   In three quadrants, the \HI\ extends further
up to about 14 kpc, whereas the N-W quadrant is dominated by an extended
filament reaching up to $8^\prime$ ($\sim$22 kpc).  Radially, the halo extends
to the end of the disk on the N-E side but stops earlier on the S-W side where
the disk is more extended.  This may be an indication that the halo is closely
connected to the inner disk of NGC\,891.

Also the kinematics of the halo is symmetric and regular in the four quadrants
up to a height of about 8 kpc (Fig.\ \ref{f_pvs_major}).  The gradient in
rotation velocity can be measured up to about 5 kpc.  Above that the rotation
velocity continues to decrease further, but the gas density is too low to
derive reliable rotation velocities
\citep{fra05}.  The gas in the halo may have  radial motions and/or
higher velocity dispersion than the disk gas, as the above model analysis indicates
(Section 4.2).  Vertical motions may also be present (as in NGC\,2403,
\citet{fra01}) but because of the inclination of the galaxy they cannot be
observed.

The halo of NGC\,891 shows individual features (streams and compact clouds)
somewhat ``separated'' either in location or kinematics from its main bulk.  
The most prominent feature is the extended filament in the N-W quadrant
 (see Figs.\ \ref{totalFig} and \ref{f_pvs_minor}).
This feature has a mass of more than 1.6
$\times$ 10$^7$ \Msol, it extends out to a projected radius of about 10 kpc
from the center of the galaxy and vertically up to 22 kpc (Figs.\
\ref{f_pvs_minor}, \ref{f_clouds}).  The full length of the filament is about
30 kpc, its velocity width is $\Delta_{\rm v} \lsim$ 100 \kms\ and its mean
velocity is close to systemic for $z\gsim$10 kpc.

A key question is whether the \HI\ halo of NGC\,891 is a diffuse medium
(filling factor $\sim$ 1) or it is entirely made of discrete {\it compact}
clouds.  A first inspection of the datacube seems to indicate that most of the
gas belongs to a smooth and coherent, differentially rotating structure.  This
picture of a ``diffuse medium'' may, however, be erroneous and be produced, at
least to some extent, by projection effects along the line of sight.  Only
clouds with anomalous velocities or located at large distances from the plane
would stand out clearly (such as those shown in Fig.\ \ref{f_clouds}).
These clouds, together with the filament on the NW quadrant may be evidence
that the halo of NGC\,891 is indeed made, at least partly, of individual gas
complexes.  These complexes would be similar to those observed in non--edge-on
galaxies like NGC\,2403 and NGC\,6946 where projection effects are less
important \citep{fra02, boo05a, boo07}.  Moreover, they would have similar
properties as the HVCs and IVCs of the Milky Way \citep{wak97} and the clouds
seen near M31 \citep{wes05}.  A structure like the long NW filament would
probably appear as a ``Complex A'' or ``Complex C'' of the HVCs to an observer
inside NGC\,891.

\subsection{Origin of the \HI\ halo}

What is the origin of the \HI\ halo of NGC\,891?
We discuss here two possibilities: the galactic fountain and the accretion
from outside. The main observational facts to be accounted for are:
1) the halo is very extended and massive (about 30\% of the
total \HI\ mass);
2) the halo kinematics is dominated by differential rotation and the rotation velocity
decreases with height from the plane;
3) a high velocity dispersion or non-circular motions (maybe inflow?) are also present;
4) the halo is structured in clouds and filaments, some of which are at very anomalous
(counter-rotating) velocities;
5) As the distance from the plane increases, the radial distribution of the
gas in the halo tends to become flatter and less concentrated to the center
than in the disk.  On the southern side, the halo is radially less extended
than the disk.

\subsubsection{Fountain}

There is considerable evidence from \ha\, radio continuum and X-ray data
pointing at a galactic fountain mechanism playing a major role in forming the
gaseous halo of NGC\,891.  The \ha\ image of NGC\,891 \citep{det90, ran90}
indicates that the star formation rate in the disk, especially on its northern
side, is very high.  There is strong radio continuum emission (thermal and
non-thermal) in the disk and also (non-thermal) in the halo, extending up to
$\sim$10 kpc from the plane (see Fig.\ \ref{f_continuum}). This extent is
close to that of the \HI\ halo and indicates the presence of magnetic fields and
relativistic electrons in correspondence of the \HI\ gas. Also, there is a clear
correlation between this radio emission and the \ha\ halo \citep{det90, ran90,
dah94}.  In particular, the northern side of the galaxy is much brighter in
both components and in the NW quadrant the radio halo has a larger scaleheight
(Fig.\ \ref{contProfs}).  A corresponding N-S asymmetry is also seen in the
\HI\ halo. Fig.\ \ref{zMom}, which shows the thickness of the \HI\ layer as a
function of position and velocity, indicates that in the inner regions of the
disk (corresponding to small $R$ and velocities furthest away from systemic),
the
\HI\ disk is thicker on the northern (i.e.\ approaching)  than on the southern side.
The northern side of the disk is where star formation and consequently the fountain are strongest.

The ionized gas in the halo of NGC\,891 has a smooth component (Diffuse
Ionized Gas) but it also shows filamentary structures
\citep{how00}.  Collimated \ha\ filaments are seen reaching up to more than 2
kpc from the plane \citep{ros04}. Moreover, the radio emission is strongly
polarized. Both these features suggest the
presence of a uniform magnetic field in the halo, indicating that the
outflowing gas has also torn the $B$-field out of the disk
\citep{dah94}. Interestingly, also the long \HI\ filament found in our data is
similarly oriented.  However, in the \ha\ images there is no trace of an
ionized counterpart of the \HI\ filament  to the detection limit.

Hot coronal gas was first revealed in NGC\,891 by ROSAT \citep{bre94}.
Recently, Chandra and XMM data have confirmed this detection and have shown
the presence of filamentary substructures in the halo component extending up
to about 5-6 kpc from the plane \citep{str04} while the soft halo component
seems to be concentrated to the inner disk \citep{tue06a}. These authors were
able to show for a sample of galaxies, including NGC\,891, that the amount of
hot gas is proportional to the mechanical feedback from supernovae, concluding
that this hot gas is almost certainly produced by a fountain.  The X-ray
emission also correlates fairly well with the \ha\ emission \citep{ros04}, and
the X-ray spectrum of the halo emission in the range 0.3$-$2 keV is fitted by
a plasma with a temperature of 0.23 keV (3.7 $\times$ 10$^6$ K).

Besides the \HI\ there are in NGC\,891 also other ``cold'' extra-planar
components such as dust and CO. The dust shows up as absorption features
against the star light up to heights of about 2 kpc \citep{how97, how00}.
Such features have masses of about 10$^5$ \Msol\ (using the galactic
dust-to-gas ratio).  This cold medium does not seem to correlate, on small
spatial scales, with the warm medium observed in \ha\ \citep{how00, ros04};
thus the two phases appear to be physically distinct.  CO emission is
also observed up to about 2 kpc from the plane of NGC\,891
\citep{gar92,sof93}. This extra-planar component of molecular gas comprises about
a 35\% of the total amount of molecular gas.

All this evidence suggests that a classical galactic fountain (ionized
outflow, cooling, and cold inflow) is active in NGC\,891 and is responsible
for much of the gaseous components, at least in the lower halo ($z\lsim5$ kpc).
These
components have a total mass of the order of a few $\times$ 10$^8$ \Msol.  Among
them, the \HI\ halo is the most massive and the most extended reaching a
distance from the plane of more than 20 kpc.  This latter may be the result of an
observational bias or the indication that a different mechanism, perhaps
accretion from outside (see below), is responsible for the formation of the
upper layers.

Galactic fountain models can easily explain the distribution of the \HI\ in
the halo of NGC\,891, with the exception of the extended filament (see below). 
Indeed, only a few percent of the energy from supernovae
would be sufficient \citep{fra06}.  However, such models fail to reproduce the
kinematics of the halo gas (both ionized and \HI). The ionized component has been
initially studied using long-slit spectroscopy \citep{ran97} and more recently
using the WIYN integral field unit \citep{hea06b}  and TAURUS spectroscopy \citep{kam07}.  The gradient in rotation
velocity derived for the ionized gas in the latter study is in good agreement
with that observed in \HI\, about $-$15 \kms kpc$^{-1}$ for $1<z<5$ kpc. 
This result indicates that the \HI\ and \ha\ phases are part, at least in the
lower halo, of the same phenomenon and, therefore, probably also have the same
origin.  A similar coupling between ionized and neutral halo
gas has also been found in the spiral galaxy NGC\,2403 \citep{fra04}.
This observed gradient in rotation velocity with $z$ is significantly higher 
than predicted by the fountain models \citep{fra06, hea06b}.  The only way to reconcile
these models with the data is to find a mechanism that would make the fountain
gas loose part of its angular momentum.  One possibility is that it is braked 
by the hot halo.  However, in this case, the angular momentum transfer to the hot
halo would speed it up very quickly \citep{fra07b}. A second, more promising
possibility is that the fountain gas interacts with low angular momentum 
material accreted from the IGM (Fraternali \& Binney, in preparation).

\subsubsection{External origin}

It is clear from the foregoing discussion that a "fountain" origin for most of
the \HI\ in the halo of NGC 891 is very likely. But a continuous supply of low
angular momentum material from outside may also be needed to account for the
observed \HI\ kinematics. The amount of accreted gas, however, can be small,
perhaps not more than 10 percent  of the gas present in the halo (Fraternali
\& Binney, in preparation).
 
Is there any direct observational evidence of such gas infall in NGC 891? 
An obvious candidate for accretion is the filament in the NW quadrant extending up
to 22 kpc from the plane.  As mentioned above, this filament has a projected
velocity close to systemic.  This may indicate either of these two possibilities: 1) the
filament belongs to the inner parts of the halo and has a very small $z$
component of the angular momentum or 2) it is located in the far outer parts
of the halo. For the filament to be produced by a galactic fountain, one would
require very high kick velocities and a very high initial kinetic energy.
Using the galactic fountain model of
\citet{fra06}, we estimate that the required kick velocities to send material
up to $z=15$ kpc range from 240 \kms\ for a starting radius of 10 kpc
to 425 \kms\ for a starting radius of $R=2$ kpc.  For a mass of the filament of
1.6$\times 10^7$ \Msol, the kinetic energies would be  $1-3 \times 10^{55}$
erg.  Therefore, to produce such a filament in one single event, hundreds of
thousands of supernovae would be required. This seems quite unlikely.  If
instead the gas complex had an external origin, its filamentary structure and
its kinematics might be more easy to explain: a cloud falling onto NGC\,891
would be stretched out and might speed up to full rotation as it
gets close to the disk.

An external origin for the filament may be as cold accretion 
from the IGM.  Or it may be through a condensation
of the hot coronal gas, possibly triggered by the galactic fountain.
Alternatively, the filament could be the result of the merging of a companion with
NGC\,891.  The stellar part of the companion could be elsewhere or too faint
to be seen and the filament would be the only prominent remnant of the
process.  In this respect we note that deep \HI\ observations are a powerful
tool to look for this kind of events and could be used to detect merging at
much larger distances than optical data.

Finally, we note that the filament roughly points at the projected position of
the companion UGC 1807 (Fig.\ \ref{f_companion}) and one may wonder whether
it could have originated from a close encounter of UGC 1807 with NGC 891.
The filament would have to be either gas pulled up from the outer
disk of NGC 891 or gas stripped from the companion.
This is not impossible, in spite of the present large separation of UGC 1807 
from NGC 891 and of its symmetrical structure and kinematics, which would seem
to exclude recent tidal encounters.
The projected distance of about 80 kpc could have been
covered by the companion with a velocity of 100$-$200 \kms\ in 8$-$4 $\times$
10$^8$ yrs.
This is sufficiently longer than the dynamical time scales in the outer
parts of the companion (2 $\times$ 10$^8$ from 100 \kms\ rotation at
4 kpc radius) to account for the absence of any sign of disturbances.
On the other hand, the filament, presumably located
in the outer parts of NGC 891, would have survived long enough to be observed now.
In this connection it is interesting to draw attention to another puzzling, long-known
structure revealed by the \HI\ observations of NGC 891:
the extended southern tail. This is difficult to maintain for a long time against the
effect of differential rotation \citep{bal80} and
the possibility of a recent origin (less than 1 $\times$
10$^9$ yrs) would therefore be quite attractive.
It is therefore interesting to ask whether a tidal
encounter with UGC 1807, as considered here for the origin of the filament,
could also be responsible for the north-south \HI\ lopsidedness of NGC 891.
Note that the mass of the companion ($\sim$10\% of that of NGC 891) is not
negligibly small.

Other indications of accretion are the counter-rotating clouds shown in Fig.\
\ref{f_clouds}.  These clouds cannot be produced by a galactic fountain
as they have an orientation of the angular momentum opposite to the
gas in the disk.  They are almost certainly located in the outer halo (one is
observed at a projected distance of 28 kpc), otherwise the drag by the
co-rotating \HI\ would have quickly inverted their motion.  Note that one of these
cloud complexes is spatially (and kinematically) very close to the filament.
The total \HI\ mass present in those accreting clouds (filament included)
is of a few 10$^7$ \Msol.  Considering a typical infalling time of the order of
10$^8$ yr, this would correspond to an accretion rate of $\approx 0.1$
\Msol yr$^{-1}$.  This latter should be considered as a lower limit to the total
accretion rate as most of the accreting material may be impossible to detect
after it has interacted with the fountain halo.

\section{Summary of results}

The deep \HI\ observations of NGC 891 reported here have shown the following:

1. The \HI\ halo is considerably more extended than
shown in previous observations.  It is detected out to a vertical distance of
about 22 kpc from the galactic plane and contains  $1.2 \times 10^9$ \Msol\, 
amounting to about 30\% of the total \HI.  The halo is not smooth. It
shows structures on various scales and there are, in particular, also \HI\
clouds with very anomalous (counter-rotating) velocities. These have masses of
about $1 \times 10^6$ \Msol.  Furthermore, there is a striking long filament
extending (in projection) up to more than 20 kpc from the disk and containing
more than 1.6 $\times 10^7$ \Msol.

2. The \HI\ halo has overall regular differential rotation but it rotates more
slowly than the disk (as already found by \citet{swa97}).  The vertical
gradient of the rotational velocity is about $-$15 \kmskpc. The shape of
the halo rotation curves does not remain constant with distance from the
plane, but its rising part becomes shallower and shallower with height.
Random motions in the halo or systematic deviations from circular motion (in-
or out-flows) may also be present.

3. A significant fraction of the \HI\ halo must come from a galactic fountain,
as the high star formation rate in the disk strongly suggests. The radio
continuum, the H$\alpha$ and the X-ray data are all corroborating evidence.
However, the kinematics of the halo suggests the need of an interaction
between the fountain gas and low angular momentum material.  Such material
may be supplied by gas accretion from the surrounding IGM.  The long \HI\
filament and the counter-rotating clouds may be direct evidence of such
accretion.

\acknowledgments

The WSRT is operated by the Netherlands Foundation for Research in Astronomy
(ASTRON) with the support from the Netherlands Foundation for Scientific
Research (NWO). This research has made use of the NASA Extragalactic Database
(NED), whose contributions to this paper are gratefully acknowledged.  The
Digitized Sky Survey was produced at the Space Telescope Science Institute
under US Government grant NAG W-2166.  We thank Richard Rand and George Heald for
providing the information about their H$\alpha$ data and George Heald for
providing his version of GALMOD.





\newpage

\begin{deluxetable}{lcc}
\tabletypesize{\scriptsize}
\tablecaption{Optical and radio parameters for NGC\,891. \label{t_parameters}}
\tablewidth{0pt}
\tablehead{
\colhead{} &
\colhead{NGC\,891} &
\colhead{ref}
}
\startdata
Morphological type      &  Sb/SBb & 1,2 \\
Radio continuum centre ($\alpha$, $\delta$ J2000)  & 2$^{\rm h}$ 22$^{\rm m}$ 33.0$^{\rm
s}$  & 5\cr
&  42$^\circ$ 20$^\prime$ 57.2$^{\prime\prime}$ & 5\\
Distance (Mpc)          & 9.5 &1 \\
$L_B$   (${L_{\odot}}_B$) & 2.6$\times10^{10}$&   3 \\
Scale length  stellar disk(kpc) &    4.4 & 4 \\
SFR (\moyr) &  3.8  & 6\\
Systemic velocity (km~s$^{-1}$)& 528 $\pm$ 2    & 5 \\
Total H{\sc\ I} mass (\Msol)  & 4.1 $\times$ 10$^9$ & 5 \\
H{\sc\ I} inclination ($^{\circ}$)        & $\gsim$89 & 5 \\
Mean P.A. (no warp) ($^{\circ}$)        & 23$^\circ$  & 5 \\
Total mass ($M_{\odot}$)& 1.4 $\times$ 10$^{11}$    &5 \\
\enddata

\tablerefs{
(1) \citet{vdk81},
(2) \citet{gar95},
(3) \citet{vau91},
(4) \citet{sha89},
(5) This work,
(6) \citet{pop04}
}
\end{deluxetable}

\clearpage

\begin{deluxetable}{lc}
\tabletypesize{\scriptsize}
\tablecaption{Observational parameters for NGC~891. \label{t_obs}}
\tablewidth{0pt}
\tablehead{
\colhead{} &
\colhead{NGC\,891}
 }
\startdata
Observation dates               &  Aug.\ - Dec.\ 2002 \\
Total length of observation (hr)   & 20 $\times$ 12 \\
Effective integration time (hr)          &  204 \\
Pointing R.A. (J2000)           &  2$^{\rm h}$ 22$^{\rm m}$ 33.76$^{\rm s}$ \\
Pointing Dec  (J2000)           &  42$^{\circ}$ 20$^\prime$ 57.1$^{\prime\prime}$ \\
Velocity centre of band (\kms)       &  800\\
Total bandwidth (MHz)           &  10 \\
Total bandwidth (\kms)  	&  $\sim$2000 \\
Number of channels in observation  &  1024 \\
Channel separation in observation (kHz)        &  9.77 \\
Number of channels of final datacube              &  224 \\
Channel separation of final datacube (kHz)        &  38.5 \\
Velocity resolution (after Hanning) (\kms)	&  16.4 \\
\enddata
\end{deluxetable}

\clearpage

\begin{deluxetable}{lccc}
\tabletypesize{\scriptsize}
\tablecaption{Parameters of the data cubes. \label{t_cubes}}
\tablewidth{0pt}
\tablehead{
\colhead{} & \colhead{Full resolution} & \colhead{Intermediate resolution} & \colhead{Low Resolution}
}
\startdata
Spatial resolution  ($^{\prime\prime}$)       & 23.4 $\times$ 16.0 & 33.2
                    $\times$ 23.9& 69.6 $\times$ 58.9 \\
P.A. of synthesized beam ($^{\circ}$)& 12.3 & 16.0 &  16.0 \\
Beam size (kpc)          & 1.08 $\times$ 0.74& 1.53 $\times$ 1.09& 3.2
                         $\times$ 2.7 \\
R.m.s.\ noise per channel (mJy beam$^{-1}$) & 0.090 & 0.10 & 0.12\\
Minimum detectable  column density (3$\sigma$;  cm$^{-2}$) & $ 13.0\times
        10^{18}$ & $6.8\times 10^{18}$ & $1.6\times 10^{18}$ \\
Minimum detectable mass (3$\sigma$; \Msol\ per resolution element)&
        $9.3\times10^4$ &$1.0\times 10^5$ &  $1.3\times 10^5$ \\
\enddata
\end{deluxetable}

\clearpage

\begin{deluxetable}{lccc}
\tabletypesize{\scriptsize}
\tablecaption{Parameters of the lagging-halo models shown in
 Figs.\ \ref{f_models_chan2} and \ref{f_models_min2}.
\label{t_models}}
\tablewidth{0pt}
\tablehead{
\colhead{Model} &
\colhead{$\Delta V_{\rm rot}/\Delta z$} &
\colhead{$\sigma_{\rm gas}$} &
\colhead{$v_{\rm rad}$} \\
\colhead{} &
\colhead{(\kms\ kpc$^{-1}$)} &
\colhead{(\kms)} &
\colhead{(\kms)} \\
}
\startdata
Lagging halo 	& $-12.0$ 	  & 8 	& 0\\
High $\sigma$ 	& $-17.4$ 	  & 25 	& 0\\
Inflow 		& $-13.0$ 	  & 8 	& $-25$\\
Shallow rise 	& $-43.0 \rightarrow -14.0$ & 20 	& $-15$\\
\enddata
\end{deluxetable}

\clearpage


\clearpage

\newpage

\newpage

\begin{figure}
\figurenum{4 (cont'd)}
\begin{center}
\epsscale{0.75}
\plotone{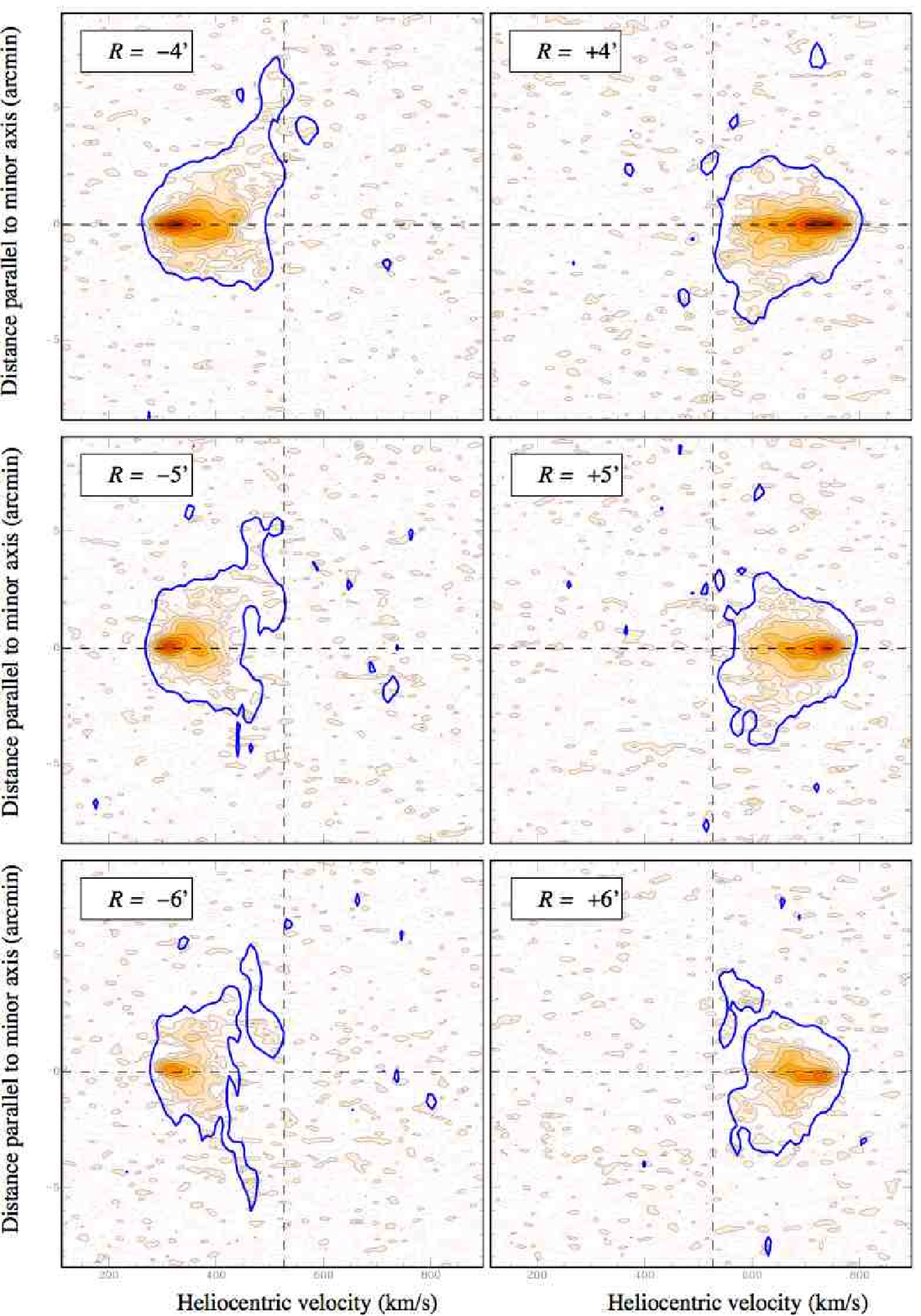}
\end{center}
\caption{Position-velocity plots perpendicular to the plane for NGC\,891  made from
the high-resolution data cube (thin lines and greyscale) and the 60-arcsec
data (thick line). Contour levels are in steps of 1.5$\sigma$ for
the high-resolution data and at 3$\sigma$ for the low-resolution data.}
\end{figure}

\newpage

\begin{figure}
\epsscale{1.0}
\plotone{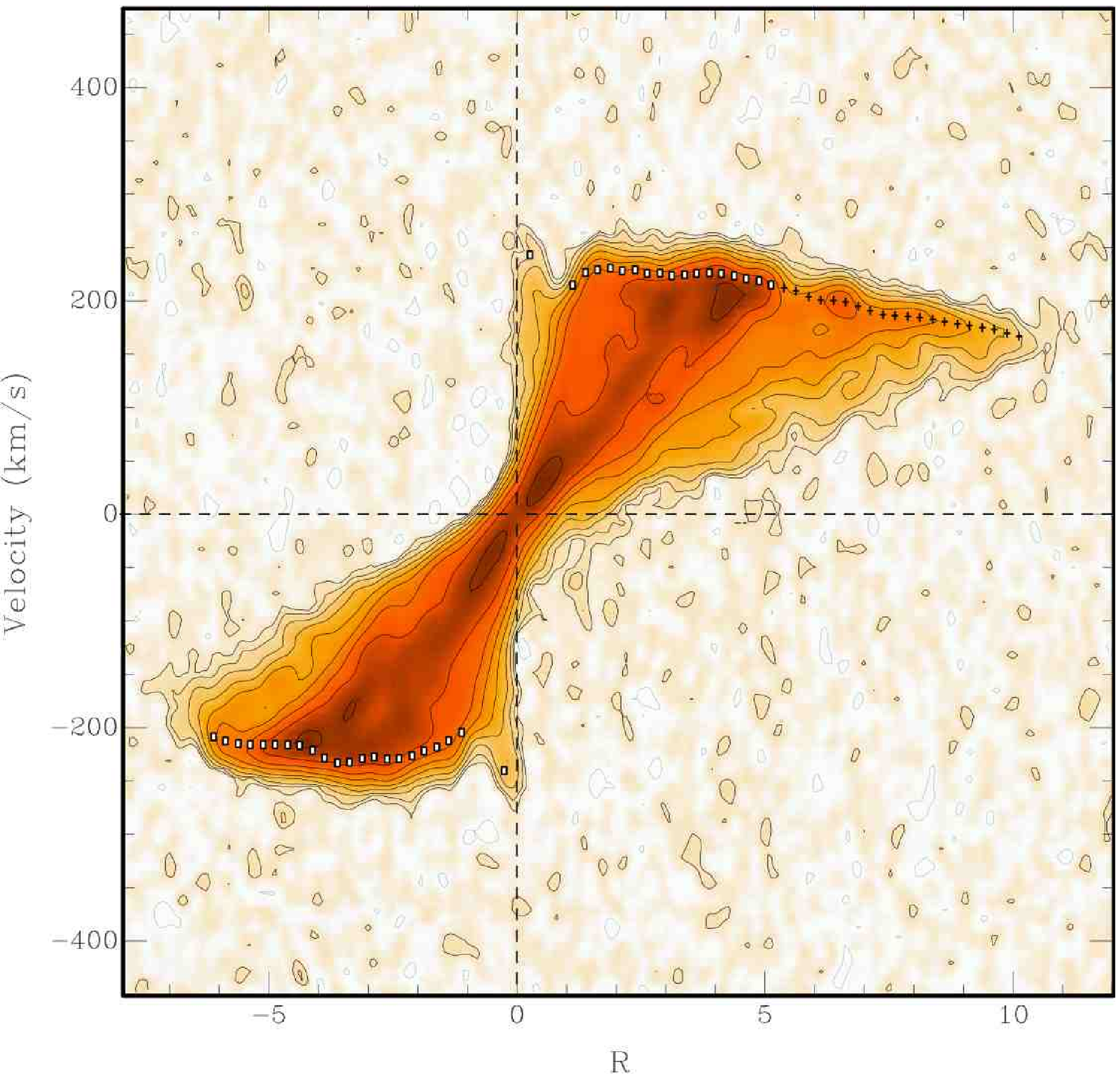}
\caption{Position-velocity diagram taken along the major axis of the
full-resolution data.  The open squares
show the rotation curve determined
for the N and S side independently, while the crosses indicate the velocities
for the SW extention of the galaxy.  Contour levels are --0.135, 0.135,
0.27, 0.54, ...  mJy beam$^{-1}$. \label{rotcurve_pv}}
\end{figure}

\newpage

\begin{figure}
\plotone{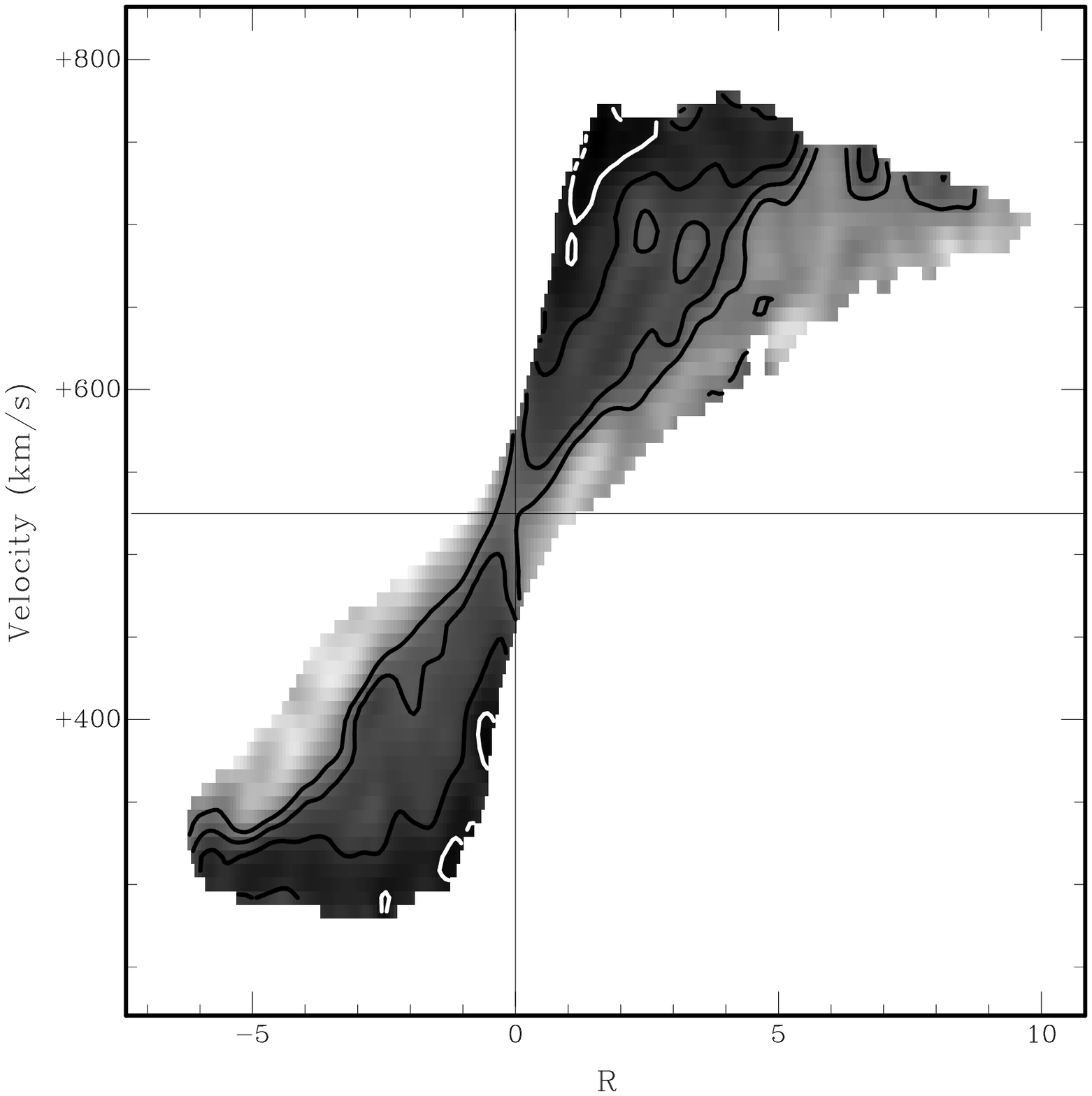}
\caption{Vertical thickness of the \HI\ emission expressed as the first moment
of the $z$ distribution at each position in the datacube with respect to the plane
of the galaxy. Contour levels are 370 (white), 465, 560 and 650 pc.
\label{zMom}
}
\end{figure}

\clearpage

\begin{figure}
\epsscale{1.0}
\plotone{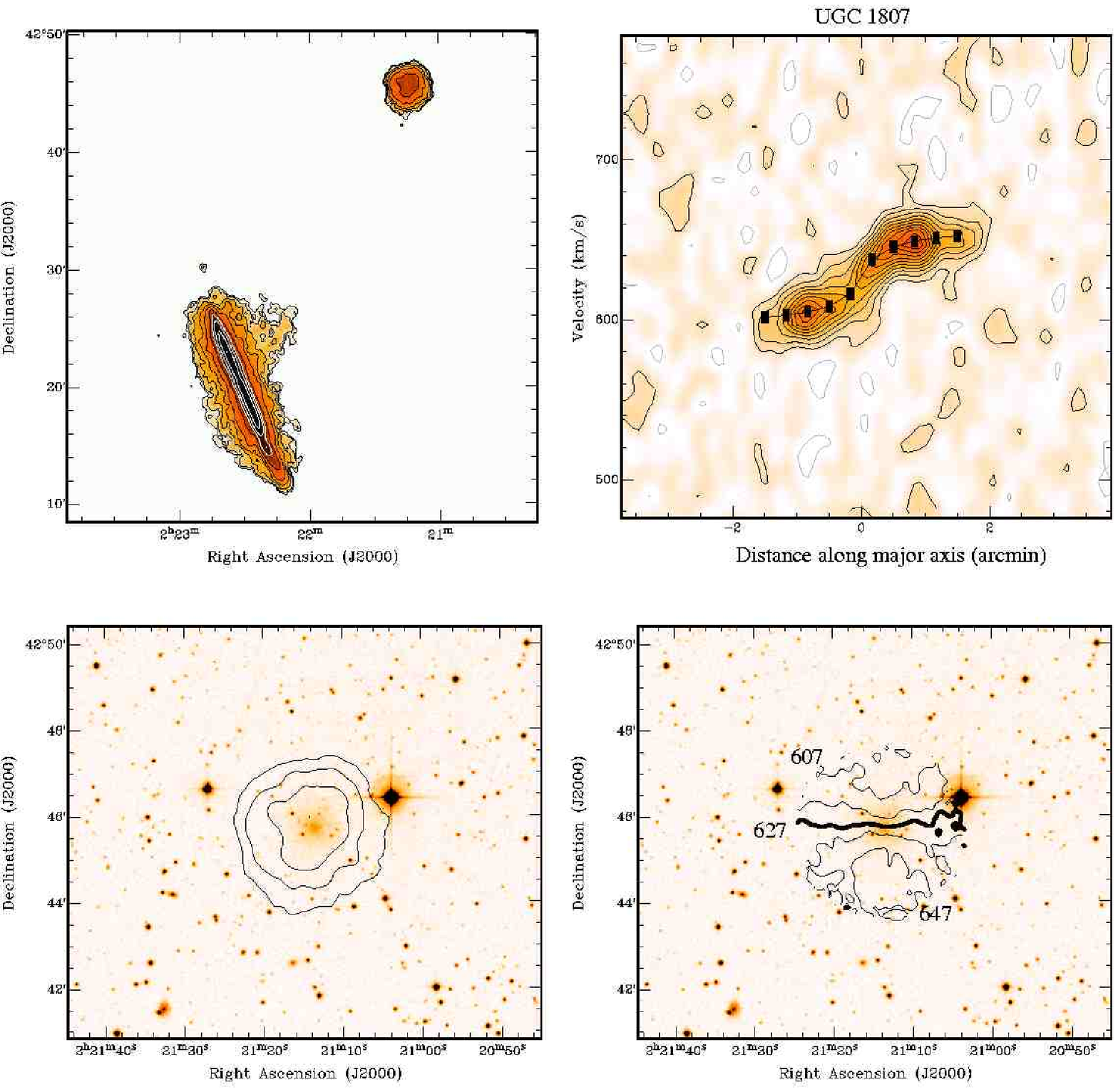}
\caption{Wide field total \HI\ map with NGC\,891 and the companion UGC\,1807
(top left);
Contours showing the total \HI\ emission of the companion, contour levels 1,
2, and $5\times10^{20}$ cm$^{-2}$ (bottom left);  velocity field of UGC 1807
(bottom right);  {\it p-V} plot of UGC\,1807 taken along the kinematical major
axis, contours  0.135, 0.27 mJy beam$^{-1}$.
\label{f_companion}}
\end{figure}

\newpage

\begin{figure}
\begin{center}
\epsscale{0.6}
\plotone{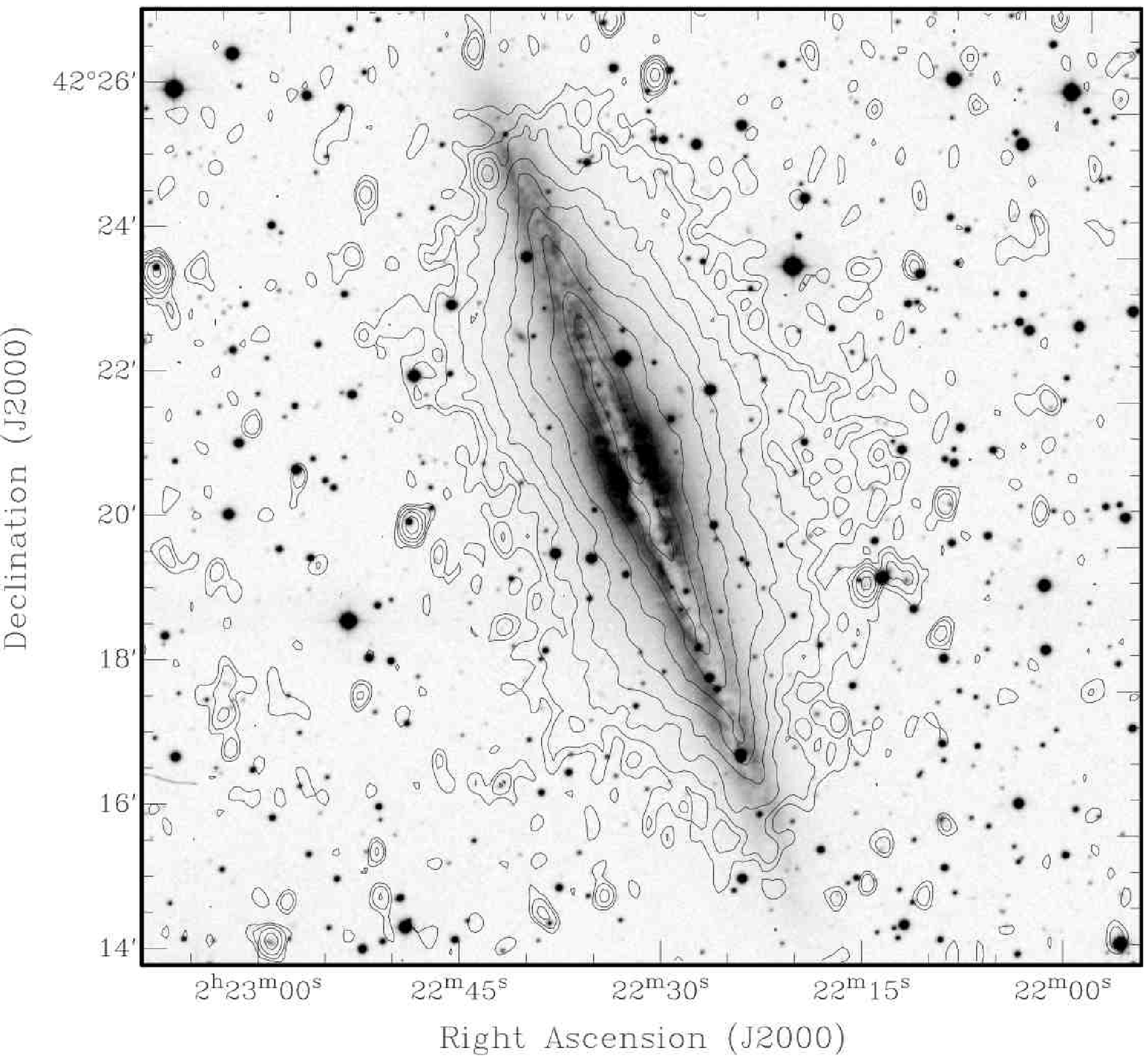}
\caption{Radio continuum contours overlaid on blue optical image taken from
the Digitized Sky Survey. Contour levels are 40, 80, 160, 320, ... $\mu$Jy beam$^{-1}$.
\label{f_continuum}}
\end{center}
\end{figure}

\begin{figure}
\begin{center}
\epsscale{0.8}
\plotone{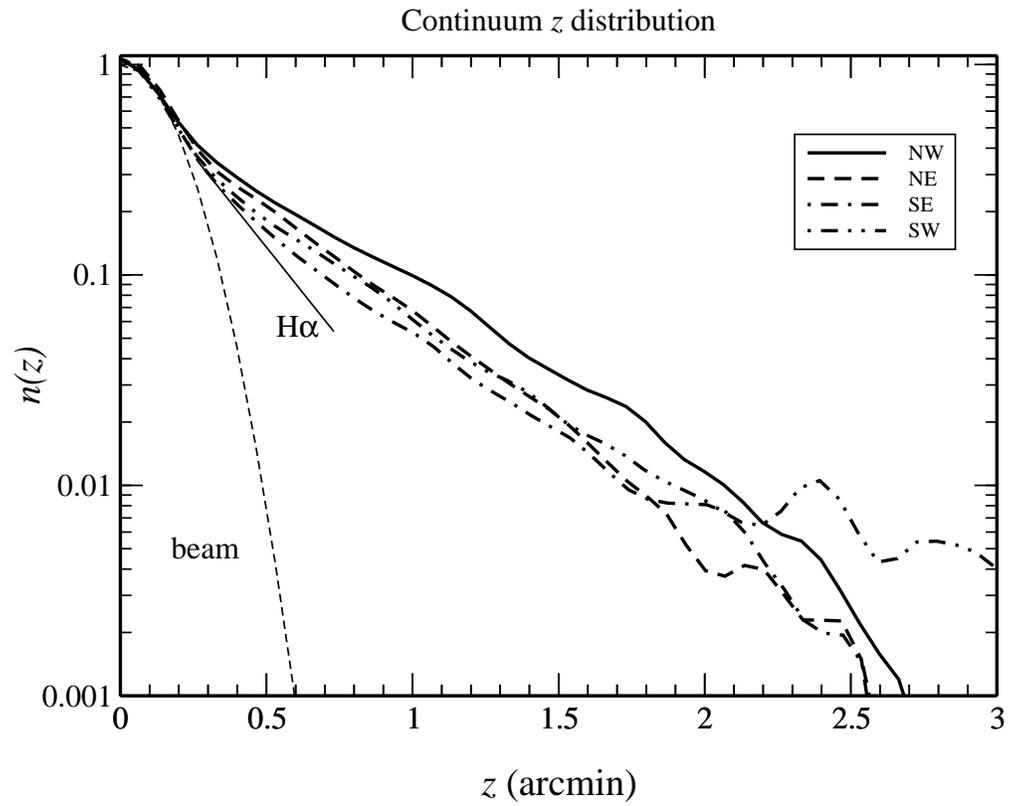}
\caption{Normalised vertical density distributions of the radio continuum
halo. The spatial resolution of the observations is indicated. The density
distribution of the H$\alpha$ emission as determined by Rand \& Heald (priv.\
comm.) is indicated.
\label{contProfs}}
\end{center}
\end{figure}

\clearpage

\begin{figure}
\plotone{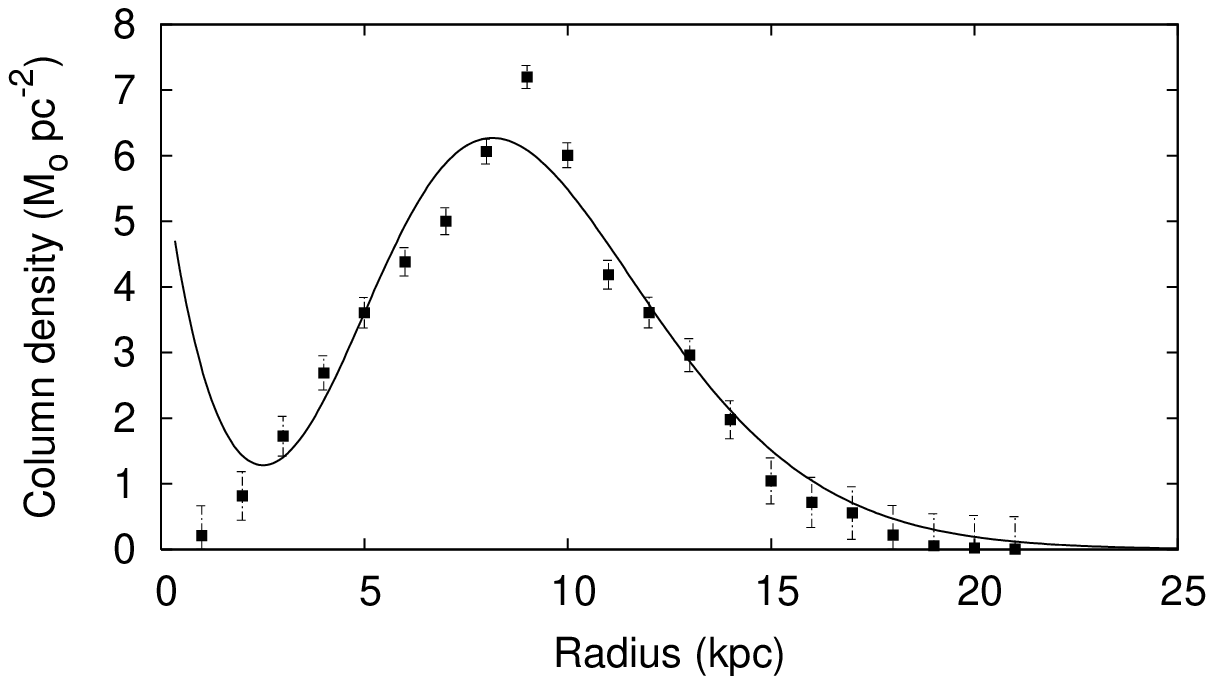}
\caption{ \HI\ radial surface density for the thin disk of NGC\,891 (squares)
overlaid with the fitting formula in eq.\ 1.
In the inner parts a compact exponential component has been added to model the
\HI\  ring.
\label{f_discdensity}}
\end{figure}

\clearpage

\begin{figure}
\plotone{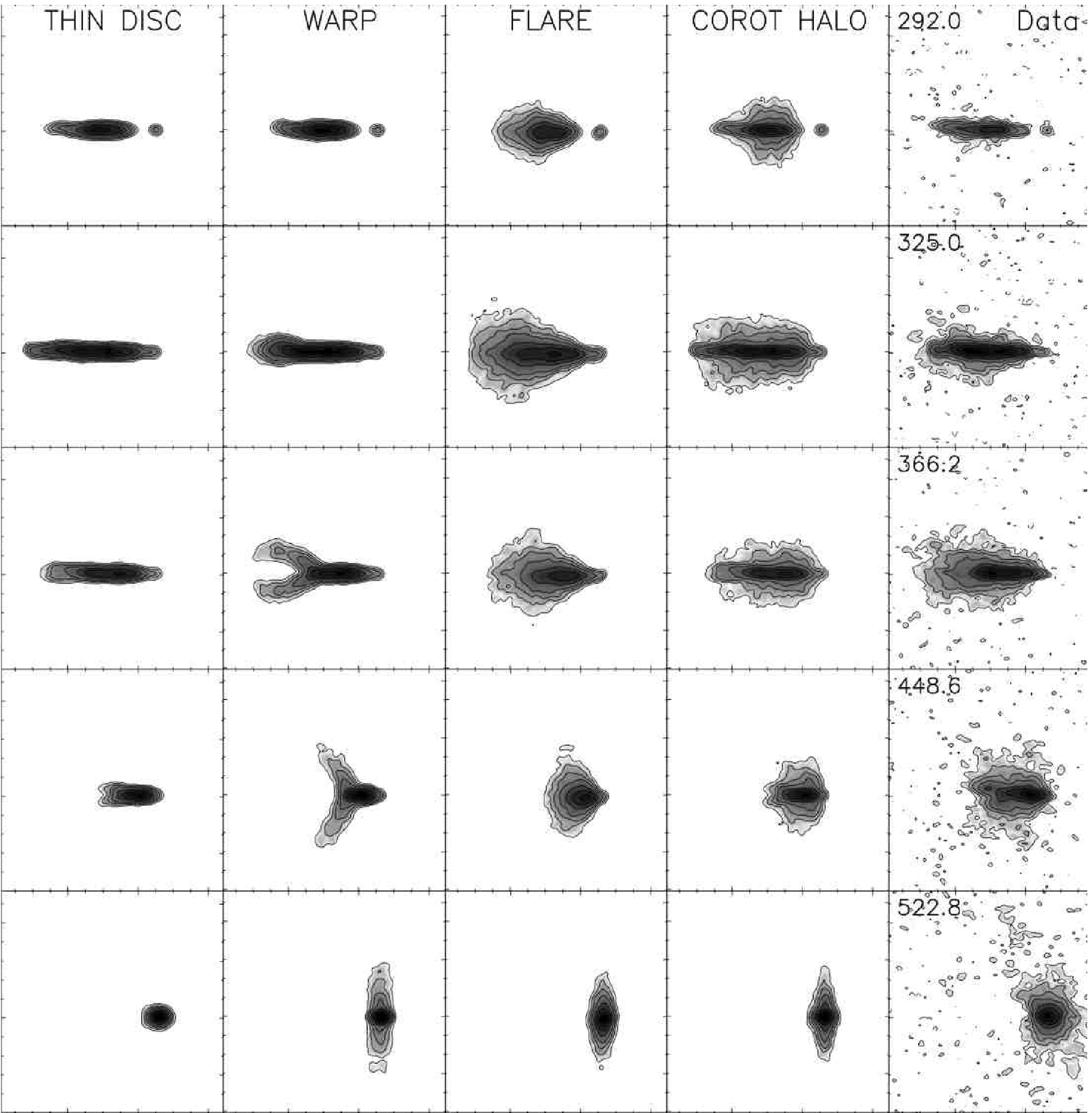}
\caption{Comparison between five representative channel maps for NGC\,891
(rightmost column) and  models.
From left to right the models are: 1) Thin disk; 2) Strong warp along the line
of sight; 3) Disk flare; 4) Disk + co-rotating halo.
Heliocentric velocities (\kms) are shown on top left corners of the data channel
maps ($v_{\rm sys}=528$ \kms).
\label{f_models_chan1}}
\end{figure}

\clearpage

\begin{figure}
\plottwo{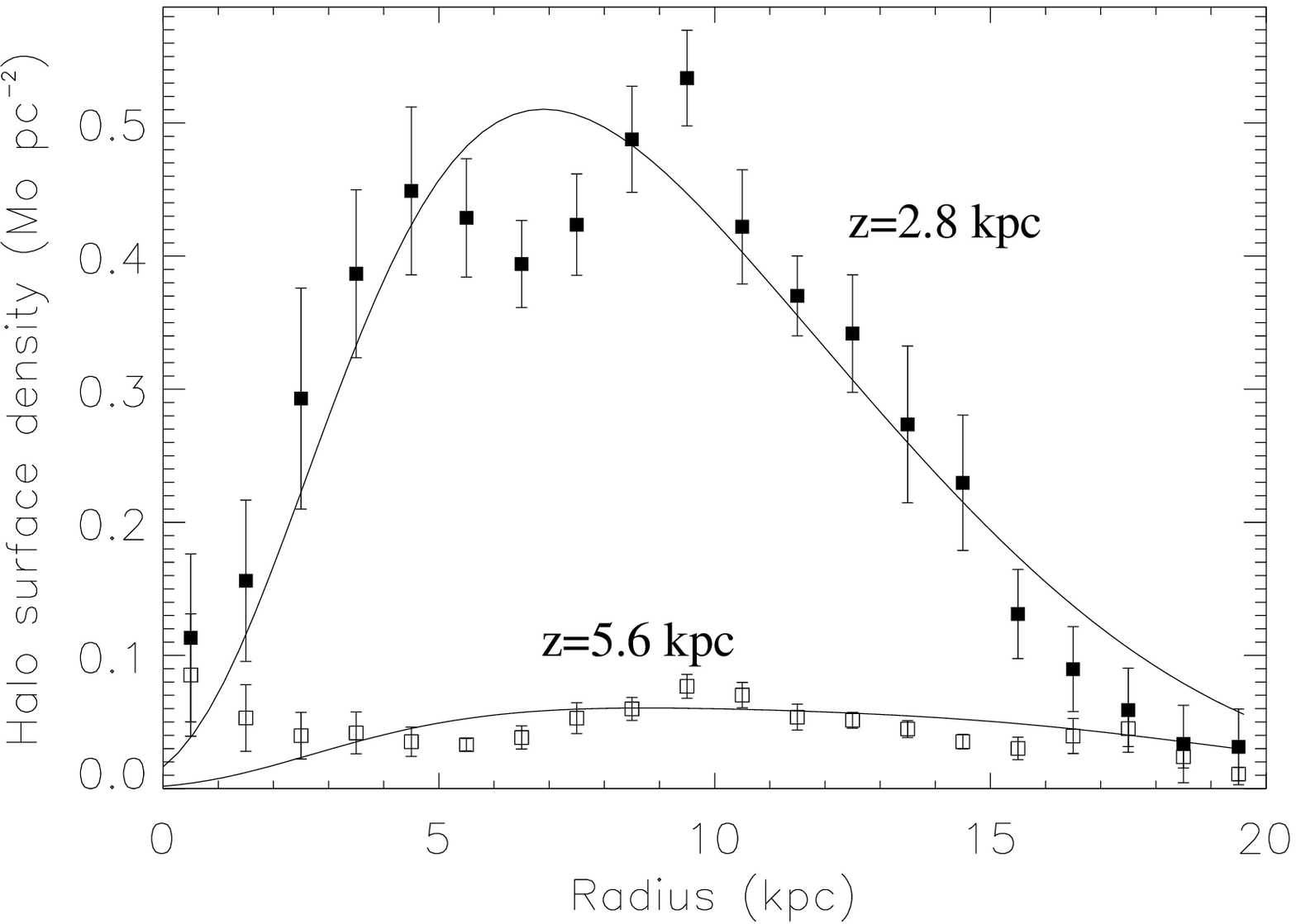}{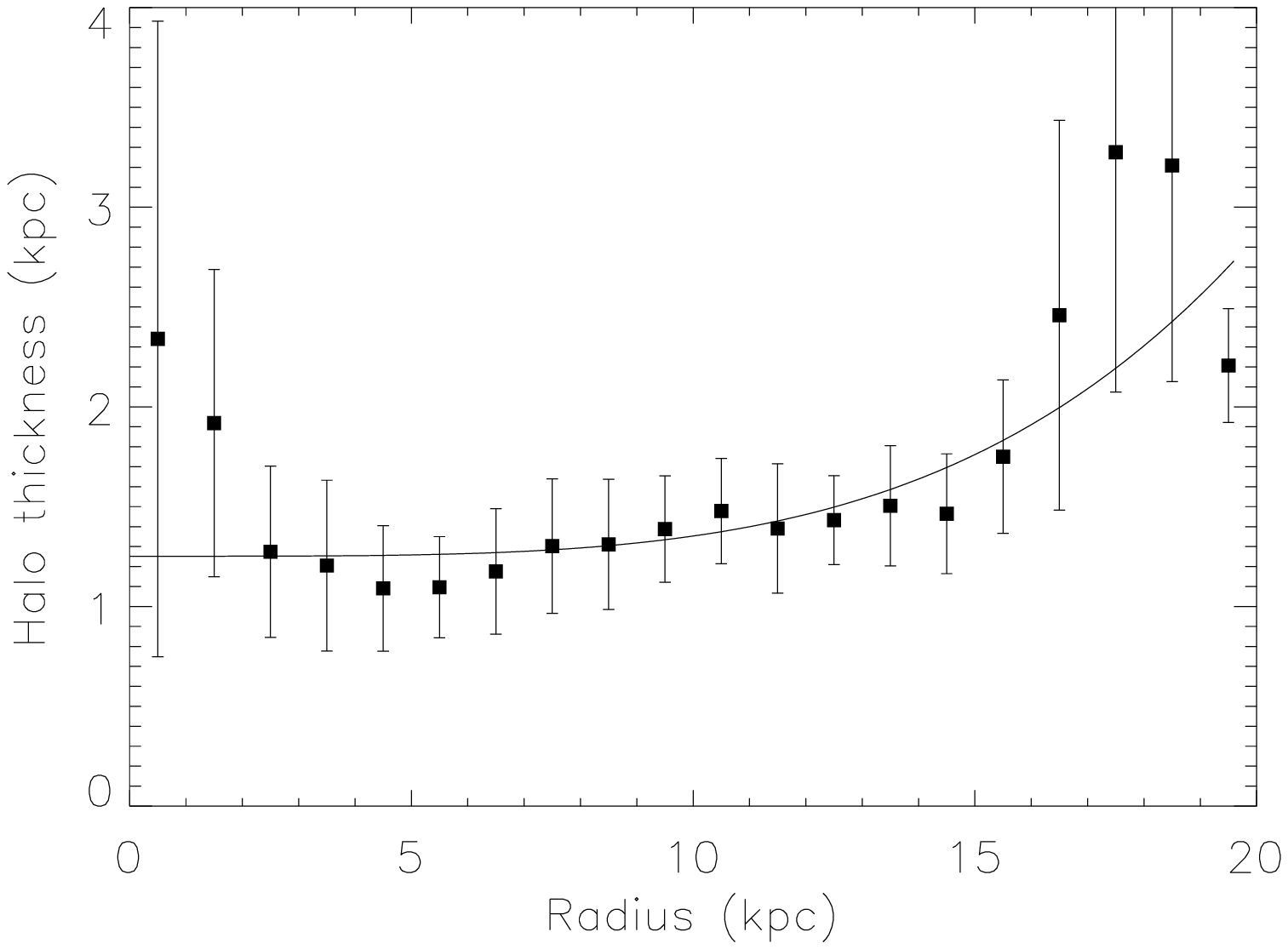}
\caption{
Gas distribution in the region of the halo.
Left panel: deprojected radial distribution at two distances
($z=2.8$ and $z=5.6$ kpc) from the plane (squares).
The lines show the results of the fit with eq.\ 3.
Right panel: scaleheight of the gaseous halo $h_{\rm halo}(R)$ (squares)
with a power law fit (solid line).
\label{f_halodensity}}
\end{figure}

\clearpage

\begin{figure}
\plotone{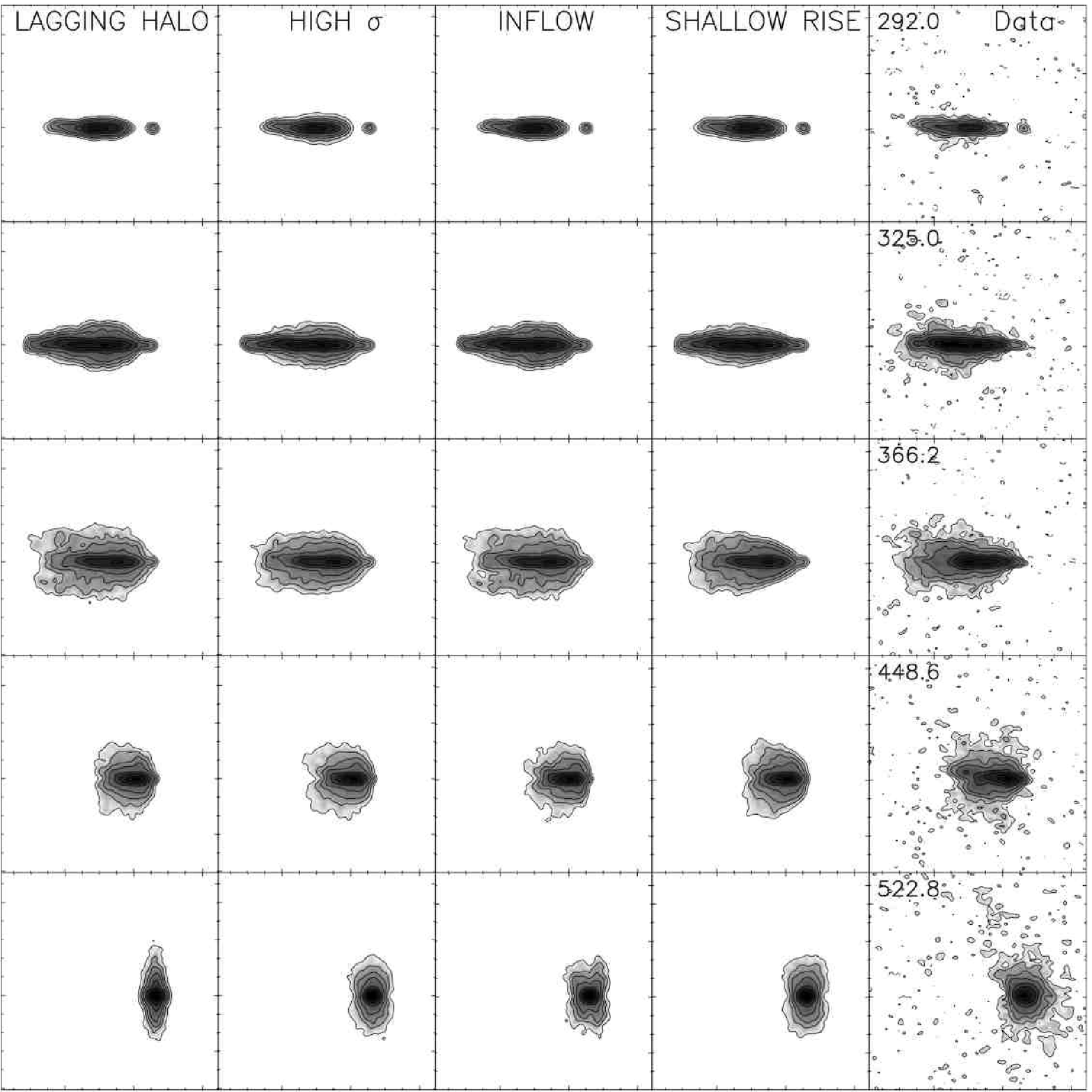}
\caption{Comparison between five representative channel maps for NGC\,891
(rightmost column) and models.
All the models have a two component (disk+halo) structure with the
halo lagging in rotation velocity with respect to the disk.
From left to right they are: 1) Lagging halo with constant gradient;
2) Lagging halo with high velocity dispersion; 3) Lagging halo with
a radial inflow motion; 4) Lagging halo with velocity gradient
increasing in the inner parts.
Heliocentric velocities (\kms) are shown in the top left corners of the 
data channel maps ($v_{\rm sys}=528$ \kms).
\label{f_models_chan2}}
\end{figure}

\clearpage

\begin{figure}
\plotone{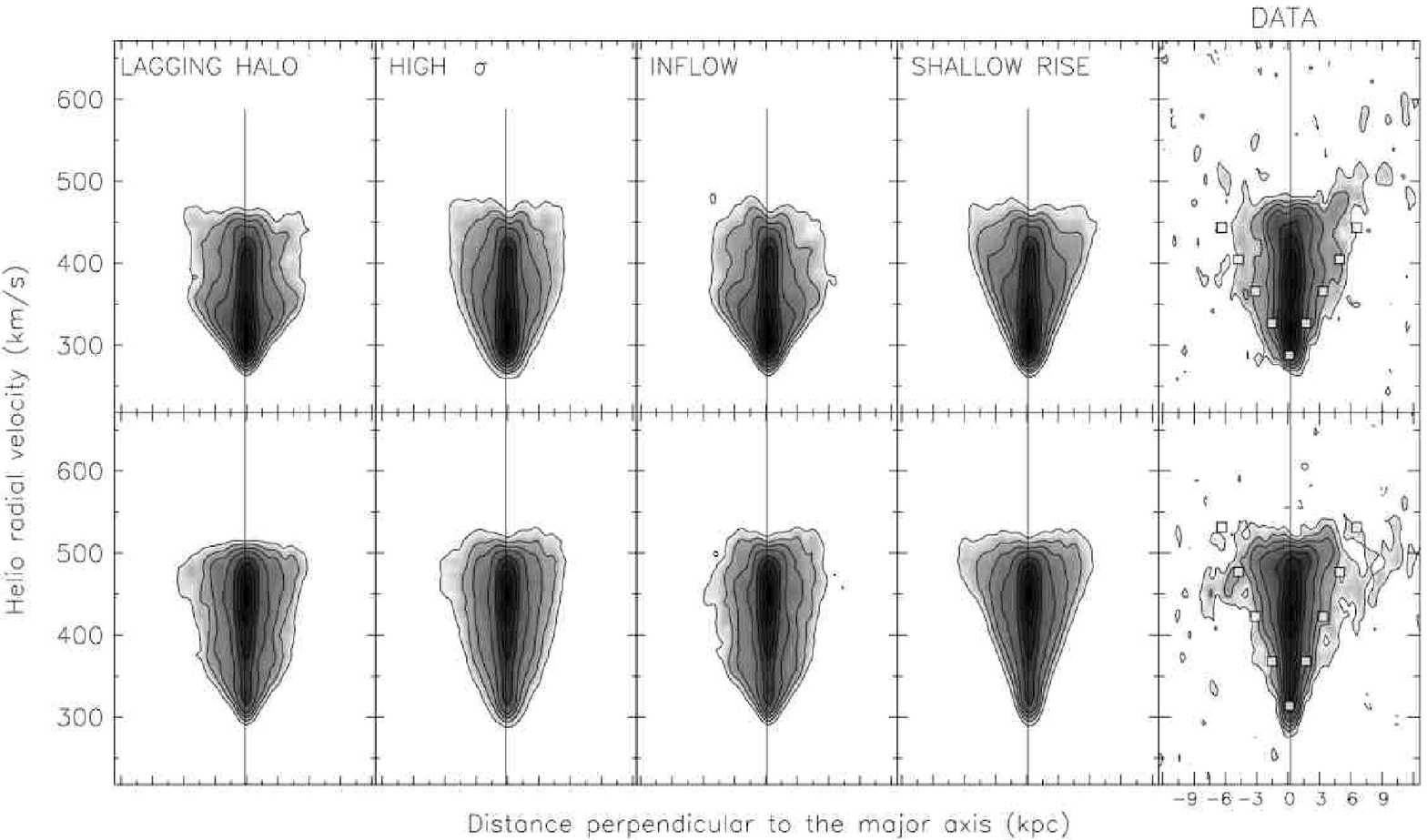}
\caption{Comparison between two position-velocity cuts perpendicular to
the plane of NGC\,891 (rightmost column) and the
models. The top row is taken 2.7$^\prime$ (7.5 kpc) North
of the centre, the bottom row 1.0 $^\prime$ (2.8 kpc)
The models are the same as in Fig.\ \ref{f_models_chan2}
\label{f_models_min2}}.
\end{figure}


\begin{figure}
\plotone{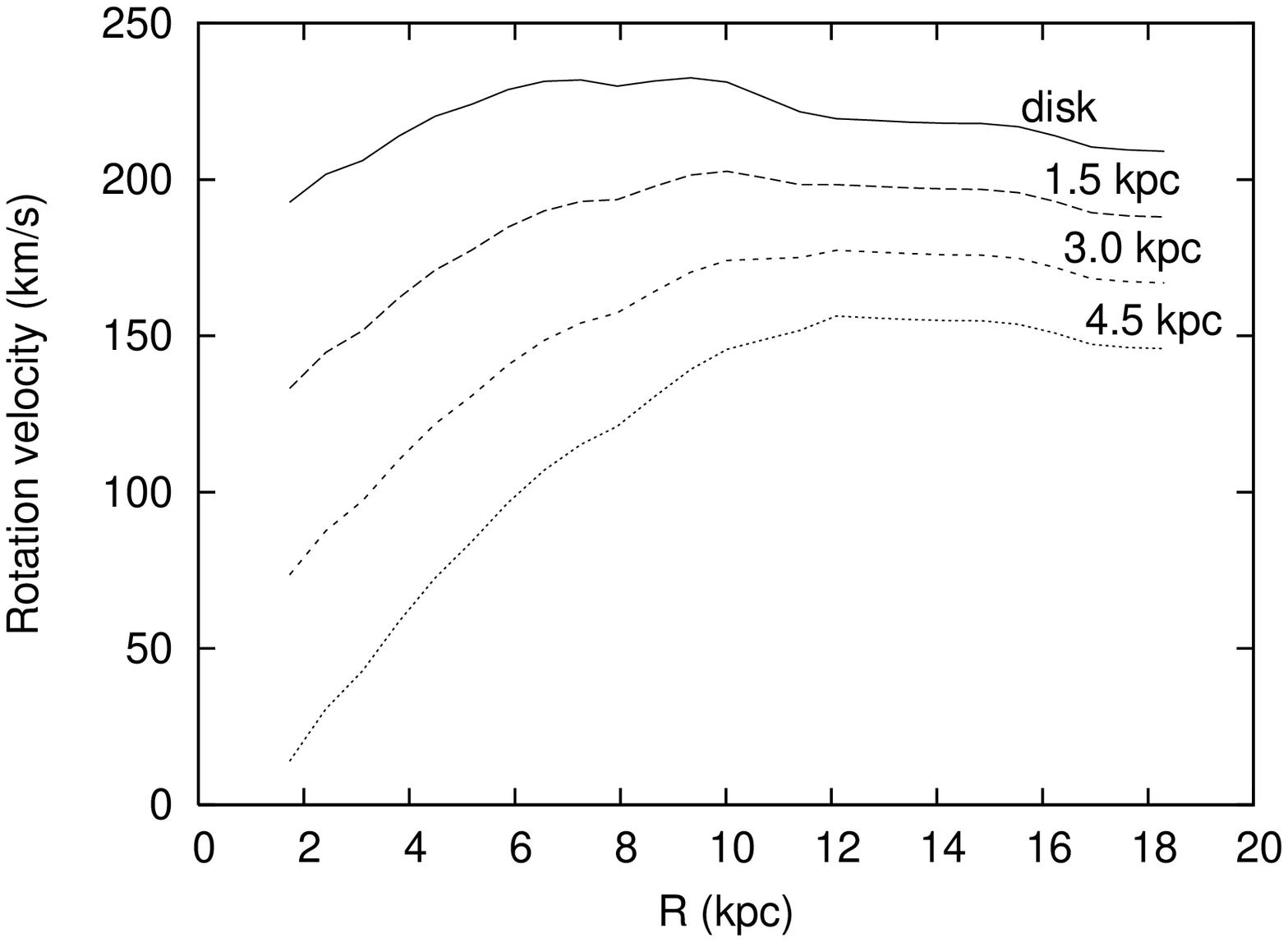}
\caption{Rotation curves  as a function of $z$ used in the
modelling. In the inner region they become shallower for increasing
$z$-distances from the plane (1.5, 3.0, 4.5 kpc).
\label{rotcurves_z}
}
\end{figure}

\newpage

\begin{figure}
\epsscale{.9}
\plotone{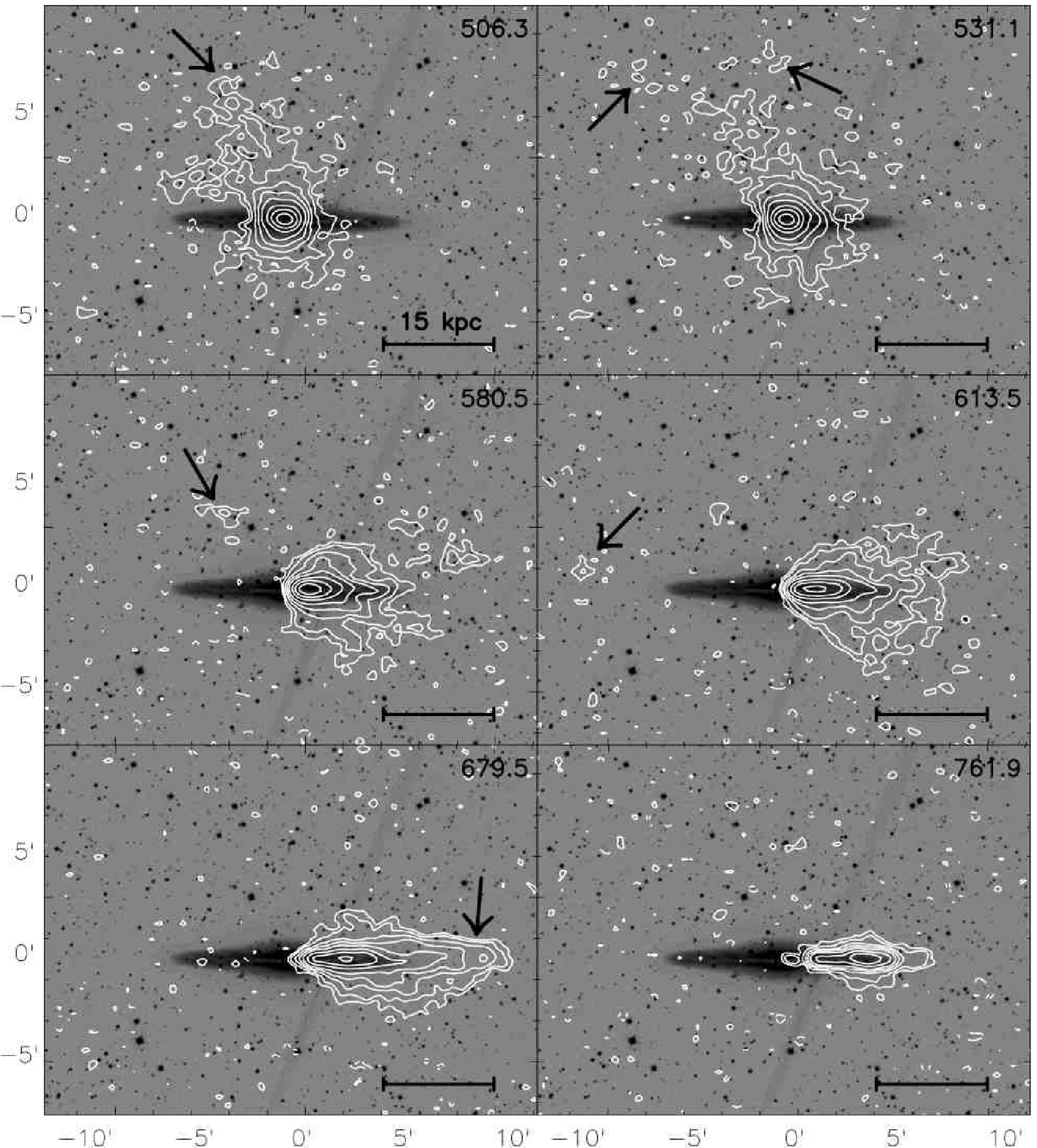}
\caption{\small Six representative channel maps for NGC\,891. 
The arrows indicate the filament
(upper panels),  two isolated ``counter-rotating'' 
clouds (middle panels), and the 
extension of the disk in the SW side (lower left panel). 
The lower right panel shows a channel map at high rotational velocity on the 
receding side. Note that both the halo gas and the extension in the disk are
absent at these velocities.
Contour levels are $-$0.4, $-$0.18, 0.18, 0.4, 1, 2, 4, 10, 20, 40 mJy/beam.}
\label{f_clouds}
\end{figure}


\end{document}